\documentclass[
  journal=pasa,
  manuscript=research-article,
  year=2024,
  volume=XX,
]{cup-journal}

\usepackage{aas-macros}

\usepackage{amsmath,amssymb,amsfonts}
\usepackage[nopatch]{microtype}
\usepackage{booktabs}

\usepackage{tabularray}

\PassOptionsToPackage{hyphens}{url}\usepackage{hyperref}
\definecolor{dodgerblue}{RGB}{30, 144, 255}
\hypersetup{colorlinks,citecolor=dodgerblue,linkcolor=blue,urlcolor=magenta}
\usepackage{xurl}  % line-breaking urls

\newcommand{\xoff}{$x_{0}$}
\newcommand{\yoff}{$y_{0}$}
\newcommand{\thetas}{$\theta_{\rm s}$}
\newcommand{\Ytot}{$Y_{\rm tot}$}
\newcommand{\aGNFW}{$\alpha$}
\newcommand{\bGNFW}{$\beta$}
\newcommand{\cGNFW}{$\gamma$}
\newcommand{\Tsz}{$T_{\rm SZ}$}
\newcommand{\thetasm}{\theta_{\rm s}}
\newcommand{\Ytotm}{Y_{\rm tot}}
\newcommand{\aGNFWm}{\alpha}
\newcommand{\bGNFWm}{\beta}
\newcommand{\cGNFWm}{\gamma}
\newcommand{\Tszm}{T_{\rm SZ}}

\newcommand{\vect}[1]{\boldsymbol{#1}}

\title{Bayesian recalibration of \emph{Planck} galaxy cluster scaling relations including relativistic Sunyaev-Zel'dovich corrections}

\author{Y. C. Perrott}
\affiliation{School of Chemical and Physical Sciences, Victoria University of Wellington, PO Box 600, Wellington 6140, New Zealand}
\email[Y. C. Perrott]{yvette.perrott@vuw.ac.nz}

 %% First letter not capped

\begin{document}

\begin{abstract}
We investigate the impact of relativistic SZ corrections on \emph{Planck} measurements of massive galaxy clusters, finding that they have a significant impact at the $\approx$\,5 -- 15\% and up to $\approx$\,3$\sigma$ level.  We investigate the possibility of constraining temperature directly from these SZ measurements but find that only weak constraints are possible for the most significant detections; for most clusters, an external temperature measurement is required to correctly measure integrated Compton-$y$.  We also investigate the impact of profile shape assumptions and find that these have a small but non-negligible impact on measured Compton-$y$, at the $\approx$\,5\% level.

Informed by the results of these investigations, we recalibrate the \emph{Planck} SZ observable-mass scaling relation, using the updated NPIPE data release and a larger sample of X-ray mass estimates.  Along with the expected change in the high-mass end of the scaling relation, which does not impact \emph{Planck} mass estimation, we also find hints of a low-mass deviation but this requires better understanding of the selection function in order to confirm. 
\end{abstract}

\section{Introduction}

Clusters of galaxies appear to delineate the upper end of the mass scale of objects in the Universe, and are powerful probes of astrophysics and cosmology.  In particular, given a sample of galaxy clusters with both well-enough calibrated masses and a well-enough understood selection function, cosmological parameters such as the matter density ($\Omega_\mathrm{m}$) and the matter fluctuation amplitude ($\sigma_8$) can be constrained in an independent and complementary way to other cosmological probes such as the primordial Cosmic Microwave Background (CMB) anisotropies and Type Ia supernova surveys (e.g.\ \citealt{2011ARA&A..49..409A}, \citealt{2016A&A...594A..24P}).  Further, while it is in principle possible to use hydrostatic masses for cosmology (e.g.\ \citealt{2017MNRAS.471.1370S}) studies based on intracluster medium (ICM) observables typically use a mass-observable scaling relation, calibrated on a subsample of well-studied objects, to translate from the observable to the mass for the bulk of the sample.  The problem of robustly calibrating the mass-observable scaling relation and understanding its intrinsic scatter has become one of the biggest challenges in cluster cosmology (e.g.\ \citealt{2019SSRv..215...25P}).

One main ICM observable is the Sunyaev-Zel'dovich (SZ) effect signal \citep{1969Ap&SS...4..301Z}, where CMB photons passing through the cluster are inverse-Compton scattered by the energetic electrons in the plasma.  This results in an overall shift in the CMB spectrum in the direction of the cluster.  The strength of the shift is typically parameterised by the Compton-$y$ parameter, defined as

\begin{equation}
y = \frac{\sigma_\mathrm{T}}{m_\mathrm{e} c^2} \int P_\mathrm{e} \mathrm{d}l,
\end{equation}
where $\sigma_\mathrm{T}$ is the Thomson cross-section, $m_\mathrm{e}$ is the electron mass, $c$ is the speed of light, $P_\mathrm{e}$ is the electron pressure and $l$ is the line of sight.  For a thermal distribution of electrons where the temperature is low enough that relativistic effects are negligible, the intensity change as a function of frequency can then be written as

\begin{equation}
\Delta I_{\nu} \propto y \frac{x^4 \exp(x)}{\left ( \exp(x)-1 \right )^{2}} \left ( x \frac{\exp(x)+1}{\exp(x)-1} - 4 \right ),
\end{equation}
\noindent where $x \equiv h\nu/k_\mathrm{B} T_{\mathrm{CMB}}$ ($\nu$ is observation frequency; $h$ is \emph{Planck}'s constant; $k_\mathrm{B}$ is the Boltzmann constant; $T_{\mathrm{CMB}}$ is the temperature of the CMB today) and $y$ is the Compton-$y$ parameter.

The largest all-sky cluster catalogue selected via the Sunyaev-Zel'dovich (SZ) effect to date was produced by the \emph{Planck} satellite (PSZ2; \citealt{2016A&A...594A..27P}).  \emph{Planck} surveyed the sky at six high-frequency bands between 100 and 857\,GHz, with angular resolution between $\approx$5 -- 10\,arcmin and used the non-relativistic thermal SZ (tSZ) effect spectrum defined above to detect clusters and constrain the integrated Compton-$y$ parameter.  X-ray-determined masses (using \emph{XMM-Newton}) for a subsample of clusters were used to constrain the mass-observable scaling relation and hence produce a catalogue of SZ masses (alongside weak lensing masses for a more limited subsample and stacked CMB lensing constraints for the cosmological analysis; \citealt{2016A&A...594A..24P}).

\begin{figure}
\includegraphics[width=\columnwidth]{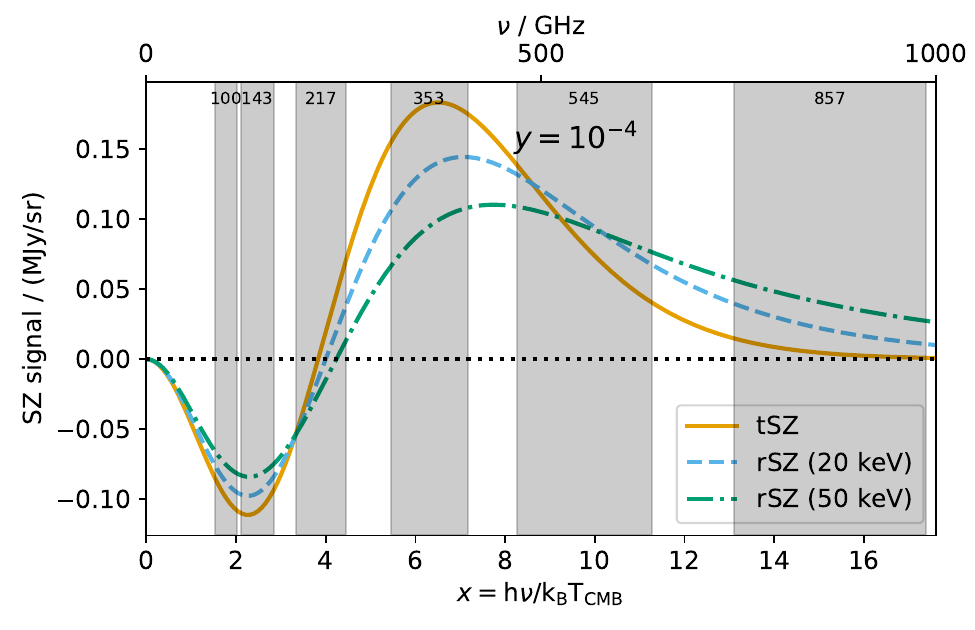}
    \caption{Orange solid line shows the SZ signal as a function of frequency assuming the non-relativistic tSZ spectrum; blue dashed and green dot-dashed show the relativistically correct rSZ spectrum with increasing temperature.  All three curves assume Compton-$y=10^{-4}$.  The grey bands show the \emph{Planck} frequency bands.}
    \label{Fi:rSZ_spectrum}
\end{figure}

In this non-relativistic limit, the tSZ spectrum does not depend on temperature.  However when electrons move at non-negligible fractions of the speed of light, higher-order, temperature-dependent corrections become necessary (\citealt{1998ApJ...499....1C}, \citealt{1998ApJ...502....7I}, \citealt{1998ApJ...508....1S}).  This is known as the relativistic SZ (rSZ) effect and becomes important at temperatures $\gtrapprox\,5$\,keV (typical for massive clusters).  The rSZ frequency spectrum is temperature-dependent (see Figure~\ref{Fi:rSZ_spectrum}): at frequencies less than $\approx$\,500\,GHz, the main change is a decrease in (absolute) amplitude of the signal with temperature, meaning that without strong constraints from higher frequency bands, there is a degeneracy between temperature and signal strength.  When the non-relativistic approximation is assumed, the effect is an underestimate of the overall cluster signal, which we refer to as the rSZ bias.  \citet{2018MNRAS.476.3360E} detected the relativistic correction to the SZ spectrum in a stacked sample of \emph{Planck} clusters at $\approx$\,$2\sigma$ level and predicted an rSZ bias of up to 14\% in the integrated Compton-$y$ parameter for the most massive clusters; \citet{2019MNRAS.483.3459R} considered the effect of the rSZ spectrum on the power spectrum of the Compton parameter and concluded it could shift the constraint on $\sigma_8$ by $\approx$\,$1\sigma$, partially alleviating the tension with $\sigma_8$ measurements from the \emph{Planck} primary CMB anisotropy data\footnote{We note that this applies when constraining cosmology directly from the Compton-$y$ power spectrum, rather than going through a mass-observable calibration as described in this paper.}.  rSZ corrections are therefore clearly becoming a non-negligible effect at the sensitivity of \emph{Planck}.

The rSZ effect is not only a source of bias: if measurable, it provides a new way to measure ICM temperatures independently to X-ray measurements.  This could shed light on the discrepancy between cluster temperatures measured with different X-ray instruments (e.g.\ \citealt{2015A&A...575A..30S}, \citealt{2024arXiv240117297M}).  In addition, observations of temperature reconstructed from the SZ effect will be weighted by the SZ signal strength and therefore pressure-weighted.  In contrast, X-ray temperature measurements are density-squared-weighted due to the well-known $n_\mathrm{e}^2$ dependence of Bremsstrahlung.  SZ temperatures will therefore be less subject to biases due to clumping and substructure (e.g.\ \citealt{2008MNRAS.386.2110K}, \citealt{2011Sci...331.1576S}, \citealt{2024arXiv240408539K}).  Comparison of SZ and X-ray measurements would therefore be a useful tool for investigating cluster thermodynamic properties and substructure.

In this paper, we investigate two main questions: (i) are \emph{Planck} cluster measurements significantly biased by ignoring the rSZ effect, and (ii) if so, is it possible to constrain cluster temperatures using \emph{Planck} data?  We also investigate the issue of pressure profile shape and how varying the profile impacts on \emph{Planck} Compton-$y$ constraints.  Finally, we produce an updated \emph{Planck} scaling relation taking into account both of these effects.

The paper is organized as follows.  In Section~\ref{S:methods_data} we outline our data and analysis methods.  In Section~\ref{S:PSZ2_comp} we compare our updated analysis to the published \emph{Planck} results.  In Section~\ref{S:sims} we explore the implications of realistic rSZ corrections to simulated \emph{Planck} data and in Section~\ref{S:profile_shapes} we explore the effect of pressure profile shape variation.  In Section~\ref{S:scaling_relations} we recalibrate the \emph{Planck} mass-observable scaling relation.  Throughout the paper, unless stated otherwise we assume a flat $\Lambda$CDM cosmology with $h=0.7$, $\Omega_\mathrm{m}=0.3$.

\section{Methods and data}\label{S:methods_data}

In this section, we outline the common methods and data that we will use throughout the paper.

\subsection{Cluster models}

\subsubsection{Observational GNFW model}\label{S:obs_GNFW}

We follow the original \emph{Planck} methodology in using a generalized Navarro-Frenk-White (GNFW, \citealt{2007ApJ...668....1N}) model for the pressure profile of the cluster gas, i.e.\
\begin{equation}\label{e:gnfw}
P_{\rm e}(r) = \frac{P_{\rm ei}}{\left(\frac{r}{r_{\rm p}}\right)^{\cGNFWm}\left[1+\left(\frac{r}{r_{\rm p}}\right)^{\aGNFWm}\right]^{(\bGNFWm-\cGNFWm)/\aGNFWm}},
\end{equation}
where $P_{\rm ei}$ is an overall pressure normalisation factor and $r_{\rm p}$ is a characteristic radius.  In the \emph{Planck} analysis, the GNFW shape parameters, \cGNFW, \aGNFW\ and \bGNFW\ describing the profile shape at radii $r \ll r_{\rm p}$, $r \approx r_{\rm p}$ and $r \gg r_{\rm p}$ respectively are set to the `universal' values derived in \citet{2010A&A...517A..92A}.  An extra parameter $c_{500}$ is also required to convert from the characteristic radius to the physically meaningful radius $r_{500}$\footnote{$r_{\Delta}$ is defined as the radius at which the enclosed mean density is $\Delta$ times the critical density at the cluster redshift.}, i.e.\ $c_{500} = r_{500}/r_{\rm p}$.  $c_{500}$ was also set to the universal value from \citet{2010A&A...517A..92A} in the \emph{Planck} analysis, so that $(\cGNFWm, \aGNFWm, \bGNFWm, c_{500})$ = (0.3081, 1.0510, 5.4905, 1.177).  We will refer to a GNFW profile with this set of parameters as the universal pressure profile (UPP).

The spherically-integrated Compton-$y$ parameter (measured in arcmin$^2$) in this model has an analytic solution when the integral is taken to infinity:
\begin{equation}\label{e:Ytot}
  \Ytotm = \frac{4\pi \sigma_{\rm T}}{m_\mathrm{e} c^2} P_{\rm ei} D_\mathrm{A} \thetasm^3 \frac{\Gamma \left ( \frac{3-\gamma}{\alpha} \right ) \Gamma \left ( \frac{\beta -3}{\alpha} \right )}{\alpha \Gamma \left ( \frac{\beta-\gamma}{\alpha} \right ) },
\end{equation}
where $\sigma_{\rm T}$ is the Thomson scattering cross-section, $m_\mathrm{e}$ is the electron rest mass, $c$ is the speed of light, $D_\mathrm{A}$ is the angular diameter distance, $\thetasm = r_{\rm p}/D_\mathrm{A}$ is the angular characteristic scale and $\Gamma$ is the gamma function.  In practice, a model cluster must be cut off at some point and in \emph{Planck} standard analysis the cut-off point is chosen to be at $5 \theta_{500}$.  With the UPP profile values this corresponds to the radius containing 96\% of \Ytot, for a spherical integral.

In the context of this model, a cluster is defined by the six parameters (\thetas, \Ytot, \cGNFW, \aGNFW, \bGNFW, $c_{500}$) plus two positional parameters (\xoff, \yoff) which we define as the offset in arcsec between the detected position and the PSZ2 catalogue position.  We assume spherical symmetry.  We refer to this model as the `observational GNFW' model since it relies entirely on observational parameters which can be constrained without knowing the cluster redshift.

\subsubsection{Physical NFW-GNFW model}\label{S:DMGNFW}

\begin{figure}
\includegraphics[width=\columnwidth]{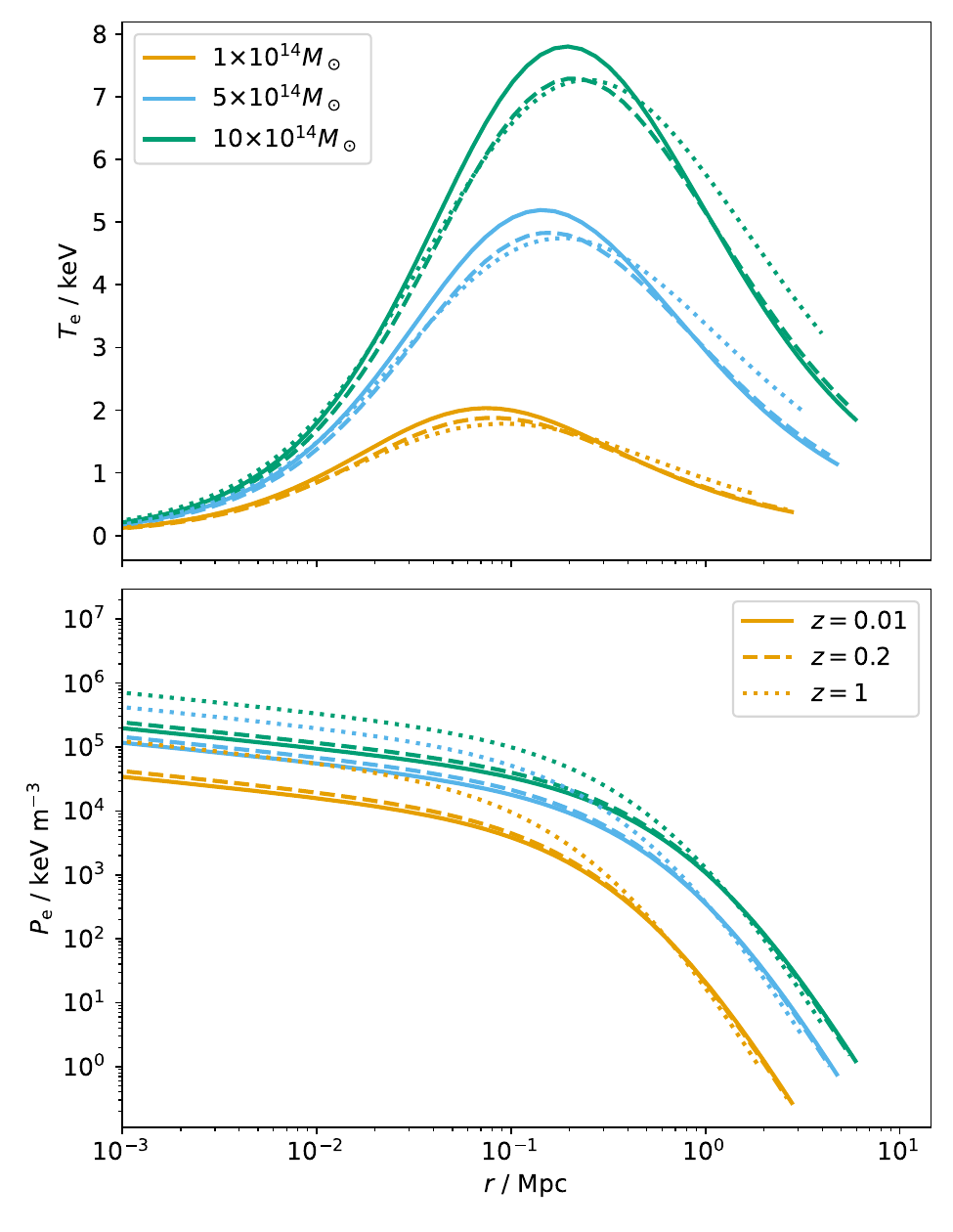}
    \caption{Example temperature (top) and pressure (bottom) profiles generated using the physical NFW-GNFW model.  In each plot, the colours indicate the $M_{200}$ mass of the simulated cluster, as given in the legend of the temperature plot.  The line styles indicate the redshift as given in the legend of the pressure plot.}
    \label{Fi:DMGNFW_demo}
\end{figure}

To generate simulations incorporating the rSZ spectrum, we will also use a physically motivated model (\citealt{2012MNRAS.423.1534O}, \citealt{2013MNRAS.430.1344O}).  This model consists of a Navarro-Frenk-White (NFW; \citealt{1997ApJ...490..493N}) density profile for the dark matter component of the cluster, i.e.\ 
\begin{equation}
\rho_{\mathrm{DM}}(r) = \frac{\rho_s}{\left ( \frac{r}{R_s} \right ) \left ( 1+\frac{r}{Rs} \right )^{2}},
\end{equation}
\noindent where $\rho_s$ is a normalization coefficient and $R_s$ is the scale radius at which $\mathrm{d}\ln \rho(r)/\mathrm{d} \ln r = -2$.  The halo concentration parameter, $c_{200} = r_{200}/R_s$ then defines $r_{200}$.  We use the \citet{2007MNRAS.381.1450N} model for $c_{200}$.  As in the observational GNFW model, the electron pressure profile is defined by equation~\ref{e:gnfw} and we assume the UPP profile parameters.  Assuming hydrostatic equilibrium, spherical symmetry and a gas mas fraction at $r_{200}$, the pressure normalization factor $P_{\rm ei}$ can be derived and therefore profiles of all thermodynamic quantities can be generated.  We use the iterative approach described in \citet{2019MNRAS.489.3135J} to refine the model and ensure it is fully self-consistent.  Some example temperature and pressure profiles generated using this model are shown in Figure~\ref{Fi:DMGNFW_demo}.

\subsubsection{Physical overdensity model}\label{S:SR_model}

Finally, to analyze \emph{Planck} data, we will also use a model based on the relationship between $M_{500}$ (total mass within $r_{500}$) and $Y_{500}$ (spherical integrated Compton-$y$ parameter within $r_{500}$).  From $M_{500}$ we can calculate $r_{500}$ given their definitions: $M_{500} \equiv 500 \rho_{\mathrm{c}}(z) \frac{4}{3} \pi r_{500}^3$.  Assuming a GNFW profile for the gas and given values for the parameters (\cGNFW, \aGNFW, \bGNFW, $c_{500}$), the parameters \thetas\ and \Ytot\ may then be calculated and the GNFW model implemented as usual.  This model can be used in two ways in a Bayesian analysis context:

\begin{enumerate}
\item Use an external prior on $M_{500}$ (i.e.\ from X-ray or lensing; in this paper we will focus on X-ray following the original \emph{Planck} methodology).  Use a non-informative prior\footnote{We use the term `non-informative prior' to refer to a prior that imposes minimal assumptions on the distribution of the parameter. This is usually either a uniform or log-uniform prior, if the order of magnitude of the parameter value is unknown.} on $Y_{500}$ to constrain the integrated Compton-$y$ value within the $r_{500}$ radius defined from $M_{500}$.  In this case we are not imposing a scaling relation between $M_{500}$ and $Y_{500}$ but results obtained in this way will be used for constraining this scaling relation.
\item Use a non-informative or mass-function-based prior on $M_{500}$; use a scaling relation between $Y_{500}$ and $M_{500}$ to sample self-consistent $Y_{500}$ and $\theta_{500}$ values.  This is a similar idea to the posterior-slicing methodology used by \emph{Planck} to post-process their \thetas-\Ytot\ constraints, but encoding it as a prior rather than a post-processing step means that uncertainties and degeneracies in all parameters can be accurately incorporated.
\end{enumerate}

We will not use (2) in this work but note its potential for future use in constraining masses from SZ data given a scaling relation.

\subsubsection{GNFW profile shape parameters}\label{S:GNFW_shape}

The GNFW gas pressure profile is a common feature of all of these models.  It is common in the SZ literature to fix the profile shape parameters to the UPP values, which were derived from X-ray observations of an X-ray-selected sample of local clusters ($z<0.2$), combined with numerical simulations to constrain the outer profile parameter (\bGNFW) which the X-ray observations did not probe.  Since the \citet{2010A&A...517A..92A} study, more progress has been made in understanding the average cluster pressure profile shape, using SZ data in combination with X-ray to probe into the outskirts, and investigations have also been made into its intrinsic scatter.

For example, \citet{2023ApJ...944..221S} performed a detailed analysis of \emph{Planck} and \emph{Bolocam} SZ effect data, in combination with \emph{Chandra} and ROSAT X-ray data, to measure the average pressure profile for a sample of clusters with redshifts ranging from 0.054 to 0.589 and masses ranging from ($3.7$ -- $22.1) \times 10^{14}$ M$_\odot$.  Dividing their sample into high-$z$, low-$z$ and relaxed subsamples, they found systematic differences between the average profiles of all three.  The intrinsic scatter was in reasonable agreement between the high-$z$ and low-$z$ samples, and lower (at low significance) for the relaxed sample in the core.  In all three subsamples, intrinsic scatter was minimum at $\approx$\,0.4 $r_{500}$ and increased to a maximum in the outskirts at $5 r_{500}$.  These results were found to be in reasonable agreement with numerical simulations as well as observational studies at similar redshifts (\citealt{2010A&A...517A..92A} (within the observed region), \citealt{2017ApJ...843...72B}, \citealt{2019A&A...621A..41G} at low-$z$; \citealt{2014ApJ...794...67M}, \citealt{2017A&A...604A.100G}, \citealt{2017ApJ...843...72B} at high-$z$), with the mean profiles estimated by the other studies typically lying within the dispersion estimated by \citet{2023ApJ...944..221S}, especially for the low-$z$ subsample.  This indicates that disagreement between studies of the average profile may just be caused by the specific sample used for analysis, and a single, fixed profile shape may not be appropriate for all the clusters in the \emph{Planck} sample.

We will use the profile shape parameters derived by \citet{2019A&A...621A..41G} in their joint X-ray (\emph{XMM-Newton}) and SZ (\emph{Planck}) analysis of clusters in the XMM Clusters Outskirts Project (X-COP) sample as a comparison profile to the UPP.  These clusters were selected from the first \emph{Planck} SZ catalogue and therefore should be representative of massive clusters observed by \emph{Planck}.  They are necessarily at low redshift ($z<0.1$) to allow them to be resolved by \emph{Planck}.  Their best-fit GNFW parameters fit to the stacked pressure profile are $(\cGNFWm, \aGNFWm, \bGNFWm, c_{500})$ = (0.43 $\pm$ 0.10, 1.33, 4.40 $\pm$ 0.41, 1.49 $\pm$ 0.30) (\aGNFW\ was kept fixed to alleviate parameter degeneracies); note that \bGNFW\ is significantly lower than the UPP value.  We will refer to a GNFW profile with shape parameters fixed to these best-fit values as the XCOP profile, and use it to demonstrate the effect of a change in \bGNFW\ and $c_{500}$ in particular.

Given the intrinsic scatter now being measured in the average pressure profile, we would like to allow for variation in the profile on an individual cluster level.  We therefore use uniform priors with ranges based on previous experimentation, taking into account the severe degeneracies between the parameters \citep{2019MNRAS.486.2116P} for our main scaling relation analysis.  The ranges chosen for these priors will be further justified in Section~\ref{S:GNFW_priors}.  When varying the pressure profile shape parameters, we will follow \citet{2019MNRAS.486.2116P} in defining the cluster cut-off radius as the radius at which a spherical integral contains $0.95 \times \Ytotm$, rather than depending on $c_{500}$.

\subsection{\textsc{PowellSnakes}}

We use the \textsc{PowellSnakes} software (\textsc{PwS}; \citealt{2009MNRAS.393..681C} and \citealt{2012MNRAS.427.1384C}) to analyze \emph{Planck} data in this paper.  \textsc{PwS} is a Bayesian analysis framework developed for detecting galaxy clusters in \emph{Planck} data, and is one of the three methods used to construct the \emph{Planck} cluster catalogue.  We choose to use it over the other, Multi-frequency Matched Filter (frequentist) methods as it fits in with our Bayesian analysis framework.  \textsc{PwS} assumes that the signal due to a cluster can be written as

\begin{equation}\label{ed:SourcesModel1}
\vect{s}(\vect{x};\vect{\Theta}) = A \vect{f}(\vect{\phi}) \vect{\tau}(\vect{x}-\vect{X};\vect{a}),
\end{equation}
where the vector $\vect{\Theta}$ contains the cluster model parameters; the vector $\vect{\tau}(\vect{x}-\vect{X};\vect{a})$ denotes the beam-convolved spatial template of the cluster at each frequency centred at the position $\vect{X}$ and characterised by the shape parameter vector $\vect{a}$; the vector $\vect{f}$ contains the emission coefficients at each frequency, which depend on the emission law parameter vector $\vect{\phi}$ of the source, and $A$ is an overall amplitude for the source at some chosen reference frequency.

\textsc{PwS} treats astronomical backgrounds as part of a generalized noise term and works on sky patches small enough to assume statistical homogeneity.  Assuming also that the background emission and instrumental noise are Gaussian random fields, there are no correlations between Fourier modes of the generalized noise and it is convenient to work in Fourier space.  Under these assumptions, it can be shown that the likelihood ratio for a single cluster can be expressed as

\begin{multline}\label{eq:LikeFilterComplete}
\ln \left[ \frac{\mathcal{L}_{H_\mathrm{s}}(\vect{\Theta})}{\mathcal{L}_{H_0}(\vect{\Theta})} \right] =
 A\mathcal{F}^{-1}
\left[ \mathcal{P}(\vect{\eta})
  \widetilde{\tau}(-\vect{\eta};\vect{a}) \right]_{\vect{X}} \\
- \tfrac{1}{2}A^2 \sum_{\vect{\eta}} \mathcal{Q}(\vect{\eta}) |\widetilde{\tau}(\vect{\eta};\vect{a})|^2,
\end{multline}
where $\mathcal{L}_{H_\mathrm{s}}$ ($\mathcal{L}_{H_0}$) is the likelihood of the hypothesis that the field contains (does not contain) a cluster; tildes denote Fourier transforms; $\mathcal{F}^{-1}[\ldots]_{\vect{x}}$ denotes the inverse Fourier transform of the quantity in brackets, evaluated at the point $\vect{x}$; and the usual mode wavenumber $\vect{k}=2\pi\vect{\eta}$.  The quantities $P$ and $Q$ are defined as

\begin{align}
\mathcal{P}({\vect{\eta}}) &\equiv \widetilde{\vect{d}}^t(\vect{\eta}) \vect{\mathcal{N}}^{-1}(\vect{\eta}) \vect{\psi}(\vect{\eta}) \\ \nonumber
\mathcal{Q}({\vect{\eta}}) &\equiv \widetilde{\vect{\psi}}^t(\vect{\eta}) \vect{\mathcal{N}}^{-1}(\vect{\eta}) \vect{\psi}(\vect{\eta}),
\end{align}
in which $\vect{d}$ is the data vector, $\vect{\mathcal{N}}(\vect{\eta})$ contains the generalized noise cross-power spectra, and the vector $\psi(\vect{\eta})$ has the components $(\vect{\psi})_\nu = \widetilde{B}_\nu(\vect{\eta}) (\vect{f})_\nu$, with $\nu$ labelling frequency channels and $\widetilde{B}_\nu$ denoting the Fourier transform of the beam in each frequency channel.  For the non-relativistic spectrum, this is extremely computationally efficient since $\mathcal{P}$ and $\mathcal{Q}$ depend only on the characteristics of the data and the signal spectrum and only need to be calculated once.  When considering the rSZ correction, $\vect{f}$ becomes a function of temperature and these quantities need to be recalculated at each likelihood calculation, introducing a small computational overhead which is acceptable when running \textsc{PwS} in targetted (rather than survey) mode.

We note that treating the astronomical backgrounds as a generalized noise term implies that we do not need a specific model to describe them, instead they are included in the cross-power spectrum which is estimated empirically from the data.  The effectiveness of this approach in separating the cluster signal from a spatially correlated dust signal is tested in Section~\ref{S:correlated_dust}.

In targetted mode, we utilize \textsc{MultiNest} \citep{2009MNRAS.398.1601F} to sample parameter posteriors, i.e. from Bayes' Theorem
\begin{equation}
\Pr(\vect{\Theta} | \vect{d}, H_\mathrm{s}) =
\frac{\Pr(\vect{d}|\,\vect{\Theta},H_\mathrm{s})\Pr(\vect{\Theta}|H_\mathrm{s})}
{\Pr(\vect{d}|H_\mathrm{s})}, \label{eq:BI_Params}
\end{equation}
where $\Pr(\vect{\Theta} | \vect{d}, H_\mathrm{s})$ is the posterior probability distribution of the model parameters $\vect{\Theta}$ given the data $\vect{d}$ and model $H_\mathrm{s}$; $\Pr(\vect{d}|\,\vect{\Theta},H_\mathrm{s}) = \mathcal{L}_{H_\mathrm{s}}(\vect{\Theta})$ is the likelihood of the data given the model and its parameters, $\Pr(\vect{\Theta}|H_\mathrm{s})$ is the prior knowledge of the parameters and $\Pr(\vect{d}|H_\mathrm{s})$ is the Bayesian evidence.  From equation~\ref{eq:LikeFilterComplete} we see that \textsc{PwS} calculates $\ln \left( \mathcal{L}_{H_\mathrm{s}}(\vect{\Theta}) / \mathcal{L}_{H_0}(\vect{\Theta}) \right )$ rather than $\mathcal{L}_{H_\mathrm{s}}(\vect{\Theta})$, however this merely introduces a constant offset and does not affect posterior or evidence evaluation.

\subsection{Posterior validation}\label{S:posterior_validation}

\begin{figure}
\includegraphics[width=\columnwidth]{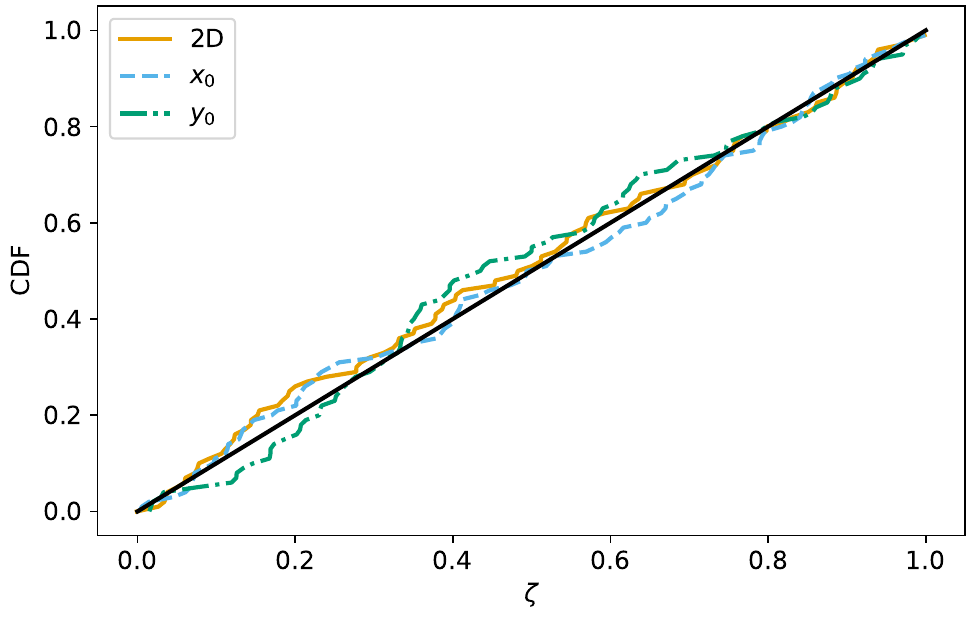}
    \caption{Posterior validation results for a high-SNR cluster simulation set, for the positional offset parameters $x_0$ and $y_0$.  The CDF of both the 2D and 1D probability mass $\zeta$ conform closely to the CDF of a uniform distribution.  This shows that the posteriors are accurate, i.e.\ the true value is contained within the 68\% contour 68\% of the time, etc.}
    \label{Fi:post_val_pos}
\end{figure}

We use the posterior validation technique from \citet{2015MNRAS.451.2610H} to test the accuracy of the posterior parameter constraints produced by \textsc{PwS} throughout the paper.  In this framework, the accuracy of a set of posterior distributions can be tested by calculating the cumulative distribution function (CDF) of the probability mass $\zeta$ contained within the highest probability density (HPD) region having the true value $x$ on its boundary.  If the posterior accurately describes the uncertainty in the parameter measurement, the CDF should follow the CDF of a uniform distribution.  This is equivalent to stating that the true value should be contained within the 95\% contour 95\% of the time; the 68\% contour 68\% of the time; and so on.  This test can be performed for any subset of the $N$ dimensions of the posterior.  Figure~\ref{Fi:post_val_pos} shows a posterior validation plot for the positional parameters in the cluster model, for a high-SNR cluster simulation providing a strict test of \textsc{PwS}'s positional accuracy.  The positional posteriors are very accurate; this result is consistent for all simulations and we will not show any further positional results in this paper.

\subsection{\emph{Planck} data and preprocessing}

We use \emph{Planck} data from the NPIPE release \citep{2020A&A...643A..42P}, incorporating the most up-to-date calibration procedures and including extra data from pointing manoeuvres, to give the best possible signal-to-noise.  We follow \citet{2014A&A...571A...9P} in calculating conversions between K$_{\rm CMB}$, MJy\,sr$^{-1}$, and SZ signal taking into account the \emph{Planck} bandpasses given in the NPIPE Reduced Instrument MOdel (RIMO), publicly available from the \emph{Planck} Legacy archive\footnote{\url{https://pla.esac.esa.int/}}, i.e.

\begin{equation}\label{eq:unit_conversion}
  \frac{{\rm d}X_i}{{\rm d}X_j} = \frac{\int {\rm d}\nu \,\tau(\nu) \left ( \frac{{\rm d}I_{\nu}}{{\rm d}X_j} \right )}{\int {\rm d}\nu \,\tau(\nu) \left ( \frac{{\rm d}I_{\nu}}{{\rm d}X_i} \right )},
  \end{equation}
where $\nu$ is frequency, $\tau(\nu)$ is the spectral transmission, $I_{\nu}$ is intensity and $X_{i/j}$ are the units of interest.  We give the conversions between K$_{\rm CMB}$, MJy\,sr$^{-1}$, and non-relativistic SZ signal in Table~\ref{T:unit_conversions}; they are consistent with the 2018 values given in the \emph{Planck} explanatory supplement\footnote{\url{https://wiki.cosmos.esa.int/planck-legacy-archive/index.php/UC_CC_Tables}}.  We assume Gaussian beams with the full-width at half maximum (FHWM) values given in the RIMO, which are also listed in Table~\ref{T:unit_conversions}.

\begin{table}
	\centering
	\caption{Unit conversions derived using the NPIPE RIMO, and effective beam full-width at half maximum (FWHM) values.}
	\label{T:unit_conversions}
	\begin{tabular}{lccc}
		\hline
		Band & MJy\,sr$^{-1}$ / K$_{\rm CMB}$ & y$_{\rm tSZ}$ / K$_{\rm CMB}$ & Beam FWHM \\
		/ GHz &  &  & / arcmin \\
		\hline
		100 & 243.8722 & -0.2480 & 9.88 \\
		143 & 371.7354 & -0.3595 & 7.18 \\
		217 & 483.4597 & 4.9487 & 4.87 \\
                353 & 287.8358 & 0.1614 & 4.65 \\
                545 & 58.1687 & 0.06921 & 4.72 \\
                857 & 2.2673 & 0.03822 & 4.39 \\
		\hline
	\end{tabular}
\end{table}

\begin{figure}
\includegraphics[width=\columnwidth]{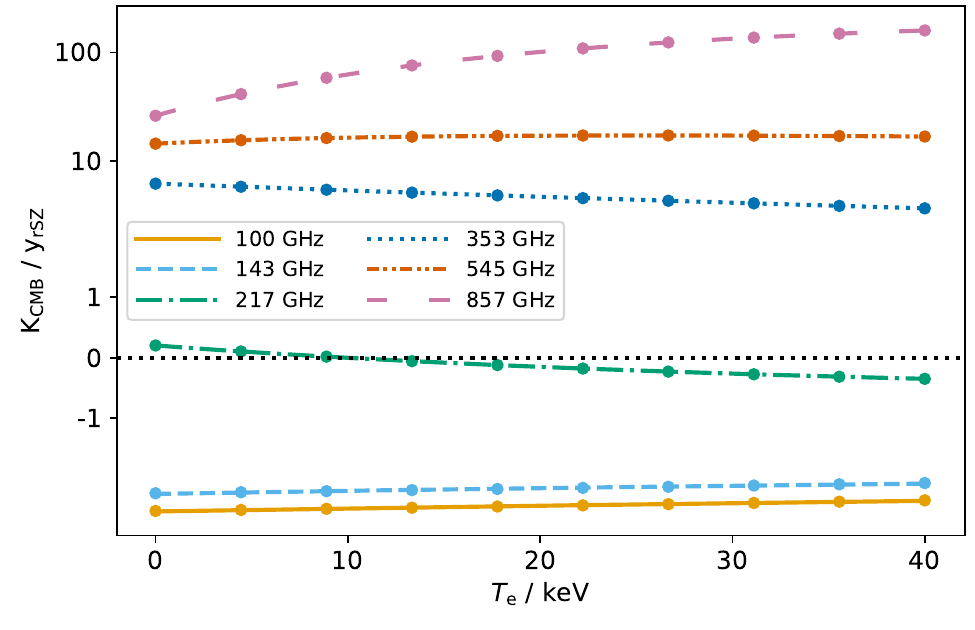}
    \caption{rSZ unit conversions as a function of temperature.  Continuous lines display conversions calculated by integrating \textsc{SZpack} calculations over the \emph{Planck} bandpasses, while dots display polynomial fits.  The $y$-axis is displayed on a `symmetric log' scale.}
    \label{Fi:unit_conversions_rSZ}
\end{figure}

The rSZ conversion factors are functions of temperature.  We calculate them for each frequency band using \textsc{SZpack} and equation~\ref{eq:unit_conversion} on a grid of temperatures ranging from 0 to 40\,keV.  For computational convenience, we fit polynomial functions to the results.  We show the calculated conversion factors with their fitted polynomials in Figure~\ref{Fi:unit_conversions_rSZ} and give the fitted polynomial coefficients in Table~\ref{T:unit_conversions_rSZ}.  We note that we fit to K$_{\rm CMB}$ / y$_{\rm rSZ}$ rather than y$_{\rm rSZ}$ / K$_{\rm CMB}$ as displayed in Table~\ref{T:unit_conversions} in the non-relativistic limit to avoid numerical instabilities near the signal null.  The polynomial fits reproduce the conversion factors to better than 0.1\%, except for the 217\,GHz frequency channel near the null (the absolute deviation is very small).

\begin{table*}
	\centering
	\caption{Fitted polynomial coefficients for the rSZ conversion factors as a function of temperature, derived using the NPIPE RIMO.  For a given frequency channel $i$, the conversion factor at a given temperature $T_{\rm e}$ may be calculated as [K$_{\rm CMB}$ / y$_{\rm rSZ}$]$_{i} = p_{0,i} + p_{1,i}T_{\rm e} + p_{2,i}T_{\rm e}^2 +p_{3,i}T_{\rm e}^3 + p_{4,i}T_{\rm e}^4 + p_{5,i}T_{\rm e}^5$.}
	\label{T:unit_conversions_rSZ}
	\begin{tabular}{lcccccc}
		\hline
		Band & $p_0$ & $p_1$ & $p_2$ & $p_3$ & $p_4$ & $p_5$ \\
                / GHz & & & & & & \\
		\hline
                100 & $-3.99$ & 0.0201 & 0 & 0 & 0 & 0 \\
                143 & $-2.74$ & 0.0134 & 0 & 0 & 0 & 0 \\
                217 & 0.202 & $-0.0239$ & 0.000394 & $-4.95 \times 10^{-6}$ & $4.33 \times 10^{-8}$ & $-1.86 \times 10^{-10}$ \\
                353 & 6.19 & $-0.0921$ & 0.000966 & $-6\times 10^{-6}$ & 0 & 0 \\
                545 & 14.5 & 0.287 & $-0.0106$ & 0.000172 & $-1.21\times 10^{-6}$ & 0 \\
                857 & 26.1 & 3.07 & 0.0958 & $-0.00435$ & $7.18\times 10^{-5}$ & $-4.73 \times 10^{-7}$ \\
		\hline
	\end{tabular}
\end{table*}

We cut square patches of side 14.7$^{\circ}$ from the NPIPE frequency maps, using \textsc{drizzlib} \citep{2012A&A...543A.103P} to accurately project from HEALpix to a WCS tangent projection.  While analyzing simulated clusters, we noted that for large clusters ($\thetasm \gtrapprox 20$\,arcmin) the patch size was slightly too small, producing an $\approx$\,5\% bias in the recovered \Ytot\ value with respect to the input value.  For these clusters, we reanalyse using double the patch side length, which eliminates the bias as long as the enlarged patch does not contain very different background characteristics such as the Galactic plane cutting through one part.

We inpaint point sources using the technique from \citet{2017PhRvD..95d3532G}.  We carry out the inpainting on all frequency channel maps if a source is detected at $>7\sigma$ in any one.  Source detections are taken from the Second \emph{Planck} Catalogue of Compact Sources (PCCS2; \citealt{2016A&A...594A..26P}) for frequency channels $<353$\,GHz and the extension to the PCCS2 using the Bayesian Extraction and Estimation Package (BeeP; \citealt{2020A&A...644A..99P}) for frequency channels $\ge 353$\,GHz.  Sources from the PCCS2E subcatalogue, from regions of the sky containing significant diffuse emission, are only selected at 100\,GHz.  We find empirically that this selection criterion includes any strong radio sources while excluding `sources' in the subcatalogue which visually appear to be knotty parts of filamentary dust emission.  True compact dusty sources from the PCCS2E appear in the BeeP catalogue at higher frequency.

\textsc{PwS} carries out an iterative background estimation pre-processing step before \textsc{MultiNEST} is run, searching for a significant cluster detection near the centre of the map and subtracting it to iteratively improve the generalized noise power spectrum estimation.  We found while analyzing simulations that this step tended to underestimate the angular size of the cluster which led to a small ($\approx$\,5\%) underestimate in \Ytot\ for high-SNR clusters.  In our updated pipeline, we alleviate this problem by running parameter estimation on the whole catalogue iteratively, at each iteration supplying an updated catalogue of cluster parameter estimates based on the previous iteration, until the parameter estimation converges.  In the background estimation step, all clusters in the field of view are subtracted based on the supplied parameters, and in the parameter estimation step all clusters except the object of interest are subtracted.  As well as removing the small \Ytot\ bias for high-SNR clusters, this improved parameter estimation for clusters close to another cluster on the map.

\subsection{X-ray data and sample}\label{S:Xray_data}

We use the X-ray hydrostatic mass estimates from \citet[hereafter L20]{2020ApJ...892..102L} to constrain our scaling relations.  The L20 sample were selected from the \emph{Planck} Early SZ (ESZ; \citealt{2011A&A...536A...8P}) catalogue and observed with \emph{XMM-Newton}.  We choose this sample because it is, to our knowledge, the largest currently available sample of X-ray masses based on a \emph{Planck}-selected sample and observed with \emph{XMM-Newton} (as in the original \emph{Planck} analysis).  We note that due to the well-known discrepancy between temperature measurements with different X-ray instruments, masses calibrated with other instruments may be systematically different (e.g.\ \citealt{2015A&A...575A..30S}).

The total L20 sample consists of 113 clusters selected from the \emph{Planck} ESZ catalogue, and further selected to have measured $r_{500}<30$\,arcmin in order to fit within the \emph{XMM-Newton} field of view.  We choose to further restrict the sample to the 103 which belong to the PSZ2 cosmological sample, which has been carefully selected to be clean of Galactic dust and point source contamination and has a well-understood selection function.  We discard two more clusters with discrepant redshifts: PSZ2~G157.43+30.34, which has a photometric redshift of $z=0.45$ in PSZ2 and L20, but now has an updated spectroscopic redshift of $z=0.402$ \citep{2018ApJ...853...36A}; and PSZ2~G055.95-34.89, with an erroneous redshift of $z=0.124$ in L20\footnote{This is possibly a transcription error; no redshifts close to this are listed for this cluster in the NASA/IPAC Extragalactic Database.} (\citealt{2015ApJ...807..178W} give a spectroscopic redshift of $z=0.2301$) leaving a total sample of 101.

\section{Initial comparison to PSZ2 cosmology catalogue}\label{S:PSZ2_comp}

\begin{figure}
\includegraphics[width=\columnwidth]{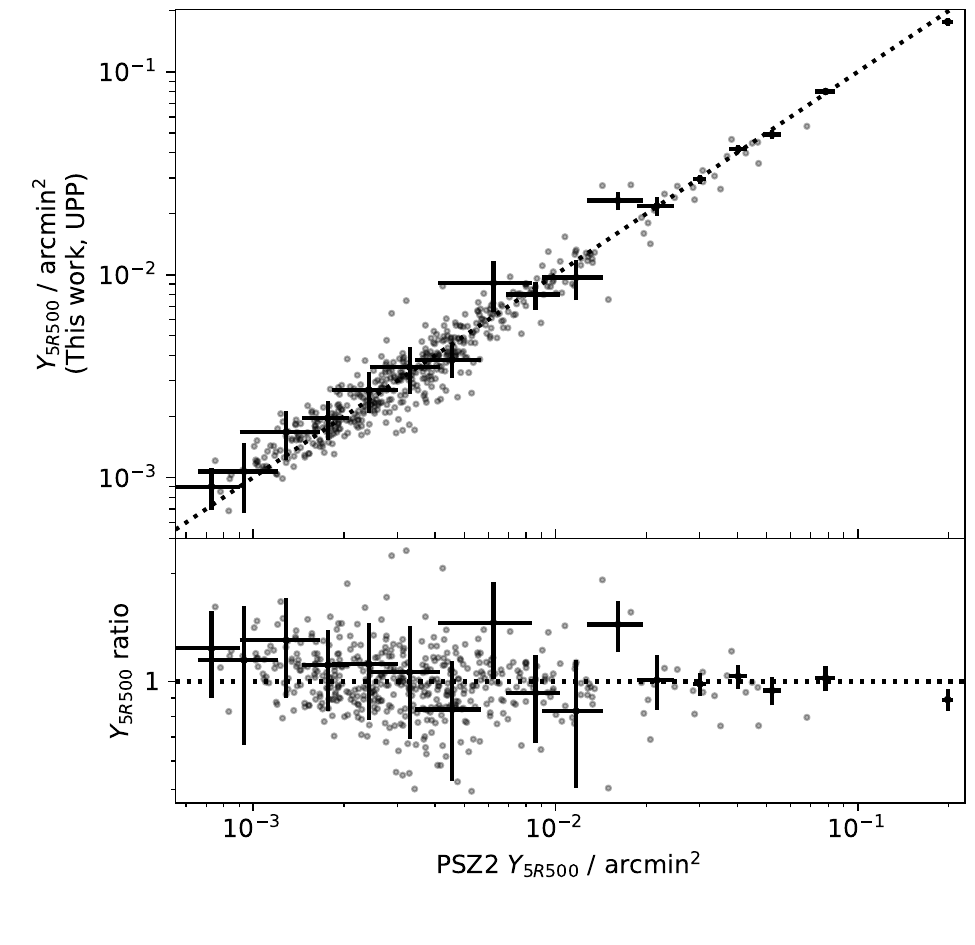}
    \caption{Comparison between $Y_{5R500}$ values from PSZ2 and this work, derived using the updated NPIPE data and preprocessing steps.  Representative error bars are shown.  The observational GNFW model with fixed UPP is assumed in both cases.  The values are compatible with no obvious biases, although some individual cluster values are offset from the one-to-one relationship.}
    \label{Fi:PSZ2_comp}
\end{figure}

As an initial comparison, we analysed all the clusters in the PSZ2 cosmology catalogue using the same methodology as the original \emph{Planck} analysis, i.e.\ assuming the observational GNFW model with the UPP, but with our updated data and pre-processing steps.  We show a comparison between the derived $Y_{5R500}$ values in Figure~\ref{Fi:PSZ2_comp}; they are compatible with no obvious systematic differences, although some individual cluster values are offset.

\begin{figure}
\includegraphics[width=\columnwidth]{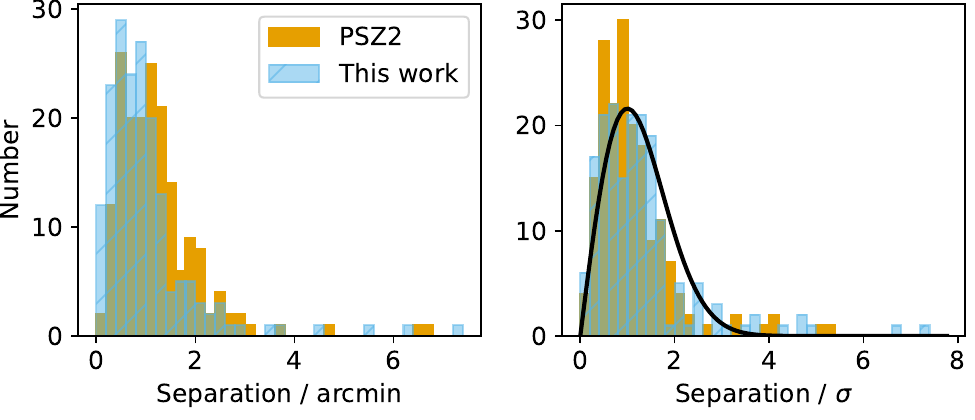}
    \caption{Offsets between matched ACT and \emph{Planck} clusters, comparing PSZ2 positions to our updated positions.  The left-hand plot shows offsets in arcmin while the right-hand plot shows offsets normalised by their error estimates, which are consistent with the Rayleigh distribution shown in black.}
    \label{Fi:ACT_offsets}
\end{figure}

We also tested the positional accuracy by matching the PSZ2 catalogue positions to the higher-angular-resolution ACT DR5 catalogue \citep{2021ApJS..253....3H} within 7\,arcmin, resulting in 178 matches.  The positional offsets are displayed in Figure~\ref{Fi:ACT_offsets}.  The comparison shows that the positions derived using our updated data and analysis method are in general closer to the ACT positions than the original PSZ2 catalogue, and therefore likely more accurate.  We also display the position offsets normalized by their errors\footnote{PSZ2 errors are 95\% confidence intervals; we divide them by $2$ to approximate the 68\% confidence interval.  Note also that the PSZ2 catalogue estimates are inhomogeneous, being drawn from \textsc{MMF1}, \textsc{MMF3} and \textsc{PwS} analysis.}; both are reasonably consistent with a Rayleigh distribution showing that the errors are appropriately estimated.  PSZ2 positional errors are around 40\% higher than our updated estimates on average.

There are two clusters with positional offsets $>5\sigma$ between ACT and our new \emph{Planck} analysis.  The first is ACO 3112 (PSZ2 G252.99-56.09/ACT-CL J0317.9-4420), which has an offset of 5.6\,arcmin or $7.3\sigma$.  The \emph{Planck} position actually matches the X-ray position to within $\approx$\,1.5\,arcmin (e.g.\ \citealt{2020ApJ...892..102L}) so we suspect a systematic error in the ACT position.  The second is ACO 2893 (PSZ2 G293.01-65.78/ACT-CL J0116.6-5046) with an offset of 7.3\,arcmin or $6.7\sigma$.  The e-ROSITA All Sky Survey catalogue \citep{2024A&A...685A.106B} contains two separate clusters with positional offsets respectively of 1.8\,arcmin from the \emph{Planck} position and 0.7\,arcmin from the ACT position; the redshifts are also different at $z\approx 0.2$ (\emph{Planck} positional match) and $z\approx 0.4$ (ACT positional match) so we conclude this is actually a spurious match.

\section{Effects of rSZ corrections to simulated \emph{Planck} data}\label{S:sims}

Before implementing a method to analyse \emph{Planck} data with the rSZ spectrum, we explored how significant the effect is and whether a resolved temperature model is necessary at \emph{Planck} angular resolution and sensitivity or whether an isothermal approximation is sufficient.

To do this, we created simulations based on some representative clusters in the PSZ2 cosmological sample.  We took the redshift and mass from the catalogue, and simulated an analogue of each cluster using the physical NFW-GNFW model described in Section~\ref{S:DMGNFW}, using the UPP GNFW shape parameters.  The model gives us radial profiles of electron pressure, density and temperature.

\begin{figure}
\includegraphics[width=\columnwidth]{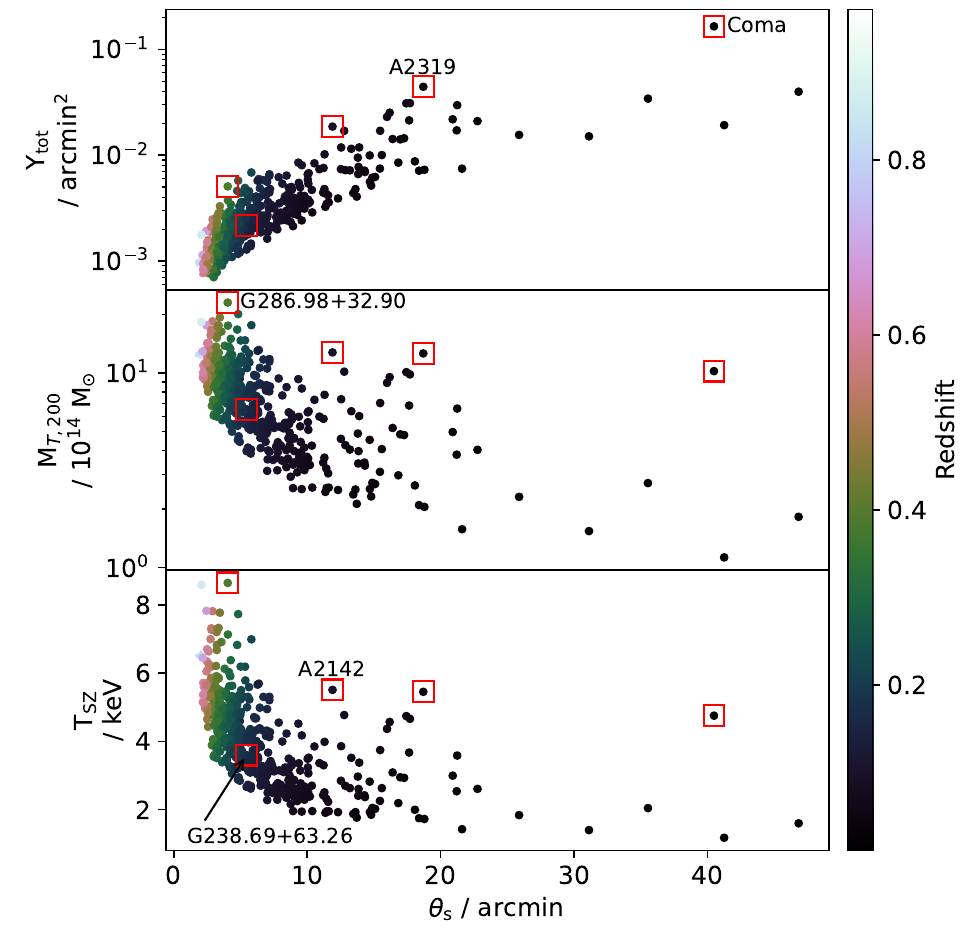}
    \caption{Illustration of the properties of the PSZ2 cosmology sample, assuming our physical model to translate from the $M_{500}$ values given in the catalogue to the properties shown.  Highlighted with boxes and labelled are the clusters we chose to simulate to test the effects of the rSZ correction.}
    \label{Fi:PSZ2_props}
\end{figure}

The rSZ correction is most likely to be significant for clusters that have very high signal-to-noise and/or are very hot; the isothermal approximation is most likely to be inadequate for clusters which, in addition, are very extended compared to the \emph{Planck} beam.  We chose clusters showing extremes in these properties as well as one which represents the median population to form the basis for our simulations.  We show these properties of the catalogue in Figure~\ref{Fi:PSZ2_props} and indicate the selected representative clusters.

%\clearpage

\subsection{Cluster simulations}

For each of the clusters selected, we simulated relativistic cluster signal maps at each \emph{Planck} frequency using the temperature-moment method implemented in \textsc{SZPack} to take into account the varying temperature both on the plane of the sky and over the line of sight, as outlined in \citealt{2013MNRAS.430.3054C}.  For these simulations we simply used the central frequency in each band, rather than taking into account the bandpass.  We approximated the \emph{Planck} beam as a Gaussian and convolved the simulated maps with a Gaussian of the appropriate width for each channel.  For comparison, we also created an isothermal cluster simulation using the pressure-weighted average temperature \Tsz, i.e.\
\begin{equation}
  \Tszm=\frac{\int P_{\rm e}(r) T_{\rm e}(r) \mathrm{d}V}{\int P_{\rm e}(r) \mathrm{d}V},
\end{equation}
where $P_{\rm e}(r)$, $T_{\rm e}(r)$ are the radial profiles produced by the model and the volume integral $\mathrm{d}V$ is taken over the entire cluster volume out to infinity.

This is the same quantity referred to as the $y$-weighted temperature in, e.g.\ \citealt{2019MNRAS.483.3459R} and \citealt{2020MNRAS.493.3274L}, however we prefer the term pressure-weighted since the Compton-$y$ parameter is defined as a line-of-sight-integrated quantity, whereas this is a 3-dimensional integral.

\begin{figure}
\includegraphics[width=\columnwidth]{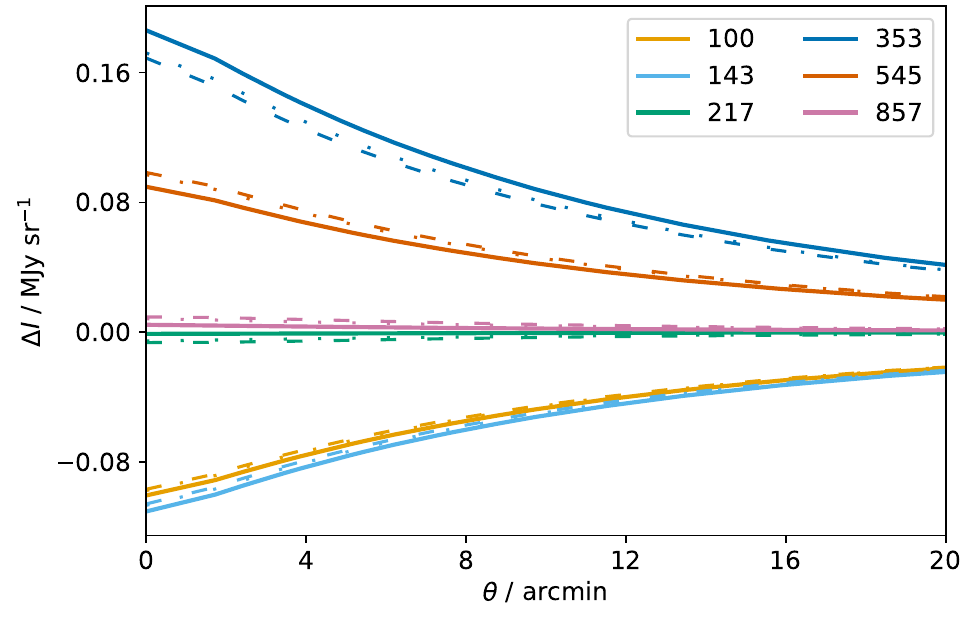}
    \caption{Map-plane signal calculated for a simulated cluster similar to Coma in each \emph{Planck} frequency channel as indicated by the colours, after line-of-sight integration and beam convolution.  The solid lines show the non-relativistic approximation; the dashed lines show the full relativistic calculation using the temperature-moment method; and the dotted lines (indistinguishable from dashed in most cases) show the relativistic calculation assuming isothermality, with the temperature equal to the pressure-weighted average over the cluster volume.}
    \label{Fi:A1656_sig}
\end{figure}

In Figure~\ref{Fi:A1656_sig} we show a comparison between the resulting simulated flux as a function of angular distance from the centre of the cluster (i.e.\ line-of-sight averaged and beam-convolved) for Coma, the most extended of our simulated clusters, when assuming (i) the non-relativistic approximation; (ii) isothermal rSZ using the pressure-weighted average temperature; and (iii) the full temperature-moment rSZ method.  It is clear that the isothermal approximation is very close to the full rSZ calculation.

We then injected the simulated clusters in real \emph{Planck} data from the NPIPE release \citep{2020A&A...643A..42P}.  We chose 100 injection positions which are at least 5 degrees away from any real cluster in the PSZ2 catalogue to avoid contamination, and at least 5 degrees away from another injection position to ensure approximate independence of the foregrounds.  Positions are also constrained to lie outside the \emph{Planck} 20 per cent Galactic plane mask.  This gave us 100 independent realizations of each cluster with realistic foreground contamination and thermal noise properties.  These 100 positions provide a fairly complete sampling of sky positions outside the Galactic mask and avoiding real clusters (it is difficult to find more than 100 positions satisfying the given conditions) but is not necessarily intended to be a truly representative set of positions such as would be required to derive a completeness function.

\subsection{Analysis of simulations}

We analysed each cluster using the observational GNFW model defined in Section~\ref{S:obs_GNFW}.  We used the non-informative priors on \xoff, \yoff, \thetas\ and \Ytot\ defined in Table~\ref{T:priors} encompassing the population shown in Figure~\ref{Fi:PSZ2_props}, and fixed the GNFW shape parameters to their UPP values.

\begin{table*}
\centering
\caption{Priors on parameters, used for analysis of both simulated and real data. $\mathcal{U}[x_{\rm min}, x_{\rm max}]$ denotes a uniform prior between $x_{\rm min}$ and $x_{\rm max}$; $\mathcal{N}(\mu,\sigma)$ denotes a Gaussian prior with mean $\mu$ and standard deviation $\sigma$ and $\delta(x)$ denotes a $\delta$-function prior fixed at $x$.  The second option for $D_\mathrm{A}^2 Y_{500}$ assumes a scaling relation with $M_{500}$, and the corresponding scatter $\sigma_{\mathrm{SR}}$ which includes both uncertainty in the mean scaling relation and intrinsic scatter.  The three options for the GNFW shape parameters are the UPP values (left), XCOP values (centre) and non-informative priors (right).  The two Gaussian options for the temperature have means corresponding to some external measurement of the temperature $\hat{T}$ (middle) and a scaling relation with some quantity $X$ which can be either mass or integrated Compton-$y$ (right), both with a fixed width of 2\,keV.}
\label{T:priors}
\begin{tblr}{
  colspec={
    l
    Q[1.5cm,halign=c]
    Q[1.5cm,halign=c]
    Q[1.5cm,halign=c]
    Q[1.5cm,halign=c]
    Q[1.5cm,halign=c]
    Q[1.5cm,halign=c]
    c
  },
  row{1}={mode=text},
  row{1}={font=\bfseries},
}
\hline
Parameter & \SetCell[c=6]{} Prior(s) & & & & & & Model \\
\hline
\xoff & \SetCell[c=6]{} $\mathcal{U}$[-600, 600] arcsec & & & & & & all \\
\yoff & \SetCell[c=6]{} $\mathcal{U}$[-600, 600] arcsec & & & & & & all \\
\hline
\thetas & \SetCell[c=6]{} $\mathcal{U}$[1.3, 60] arcmin & & & & & & Obs. GNFW \\
$\log(\Ytotm/\mathrm{arcmin}^2)$ & \SetCell[c=6]{} $\mathcal{U} \left[ \log(5\times 10^{-4}), \log(5\times 10^{-1}) \right]$ & & & & & & Obs. GNFW\\
\hline
$M_{500}$ & \SetCell[c=3]{} $\mathcal{N}(\hat{M_{500}},\sigma_{M_{500}})$ M$_{\odot}$ & & & \SetCell[c=3]{} $\log(M_{500}/\mathrm{M}_{\odot})\sim \mathcal{U}[14,15.3]$ & & & Physical overdensity \\
$D_\mathrm{A}^2 Y_{500}$ & \SetCell[c=3]{} $\log(D_\mathrm{A}^2 Y_{500}/\mathrm{Mpc}^2)\sim\mathcal{U}[-7, -3]$ & & & \SetCell[c=3]{} $\mathcal{N}(A E(z)^C M_{500}^B, \sigma_{\mathrm{SR}})$ Mpc$^2$ & & & Physical overdensity \\
\hline
\cGNFW & \SetCell[c=2]{} $\delta(0.3081)$ & & \SetCell[c=2]{} $\delta(0.43)$ & & \SetCell[c=2]{} $\delta(0.43)$ & & all\\
\aGNFW & \SetCell[c=2]{} $\delta(1.0510)$ & & \SetCell[c=2]{} $\delta(1.33)$ & & \SetCell[c=2]{} $\mathcal{U}$[0.1, 3.5] & & all\\
\bGNFW & \SetCell[c=2]{} $\delta(5.4905)$ & & \SetCell[c=2]{} $\delta(4.40)$ & & \SetCell[c=2]{} $\mathcal{U}$[3.5, 7.5] & & all\\
$c_{500}$ & \SetCell[c=2]{} $\delta(1.177)$ & & \SetCell[c=2]{} $\delta(1.49)$ & & \SetCell[c=2]{} $\mathcal{U}$[0.6, 2.4] & & all\\
\hline
\Tsz & \SetCell[c=2]{} $\mathcal{U}$[0, 40] keV & & \SetCell[c=2]{} $\mathcal{N}(\hat{T}, 2)$ keV & & \SetCell[c=2]{} $\mathcal{N}(A E(z)^C X^B, 2)$ keV & & all\\
\hline
\end{tblr}

\end{table*}

We analysed the rSZ simulations first with the non-relativistic SZ calculation, and secondly with an isothermal rSZ model, adding another parameter \Tsz\ on which we trial 3 priors: (i) a uniform prior from 0 to 40\,keV, (ii) a Gaussian prior centred on the true pressure-weighted average temperature, and (iii) a Gaussian prior linked to \Ytot\ via a scaling relation.  For the latter, we generated temperature and pressure profiles using the physical NFW-GNFW model (as demonstrated in Figure~\ref{Fi:DMGNFW_demo}) for clusters with a range of masses and redshifts, and integrated the profiles to calculate the pressure-weighted \Tsz.  We fit a scaling relation to the results of form $\Tszm=A E(z)^C (\Ytotm D_\mathrm{A}^2)^B$, and obtained $(A, B, C)=(156.65,0.39,0.26)$ for \Tsz\ measured in keV and $\Ytotm D_\mathrm{A}^2$ measured in Mpc$^2$.

\begin{figure*}
\includegraphics[width=\linewidth]{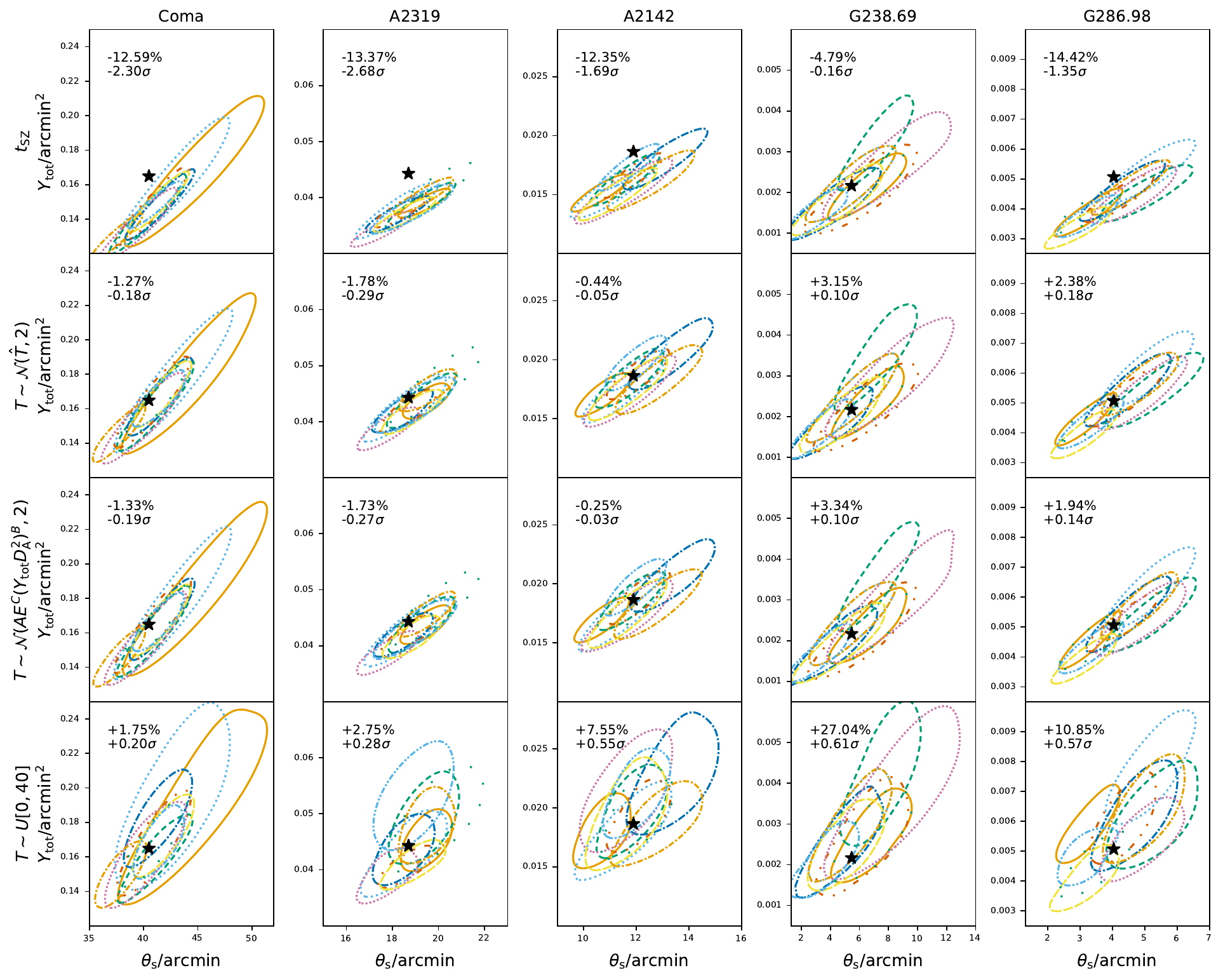}
    \caption{Posterior constraints on the cluster \thetas\ and \Ytot\ model parameters derived from analysing the full relativistic simulations with the non-relativistic approximation to the SZ signal (top row) and isothermal relativistic signal (other rows; prior on \Tsz\ given on axis).  For a given cluster, all plots have the same axis ranges and the true value is marked with a black star.  The simulation set for each cluster consists of 100 realizations where the same cluster model is input into different regions in the real \emph{Planck} sky maps.  In these plots, a random selection of ten realizations has been chosen for each cluster and their 68\% posterior probability contours shown with different colours/line styles in each plot, to display the variation due to thermal and foreground noise.  Average 1D \Ytot\ bias values for all 100 realizations in each simulation set are given on the top left-hand corner of each figure, where a negative bias value indicates that the recovered \Ytot\ is biased down with respect to the input value.}
    \label{Fi:sim_posteriors}
\end{figure*}

The resulting posterior constraints on \thetas\ and \Ytot\ are shown in Figure~\ref{Fi:sim_posteriors} for each simulated cluster and each \Tsz\ prior, for a selection of 10 out of the 100 realizations.  We also show the mean bias with respect to the true value in the 1D marginal \Ytot\ constraint as a percentage and a $\sigma$ level, averaged over all 100 realizations.  It is clear that the relativistic correction does have a significant effect, with \Ytot\ biased down by $\approx$\,5 -- 15\% when the non-relativistic approximation is used for analysis.  In the higher SNR cases, it is also significant at the $\approx$\,2 -- 3$\sigma$ level.  In lower SNR cases, the bias is not significant at the individual cluster level but will be significant over the whole \emph{Planck} sample.

\begin{figure*}
\includegraphics[width=\linewidth]{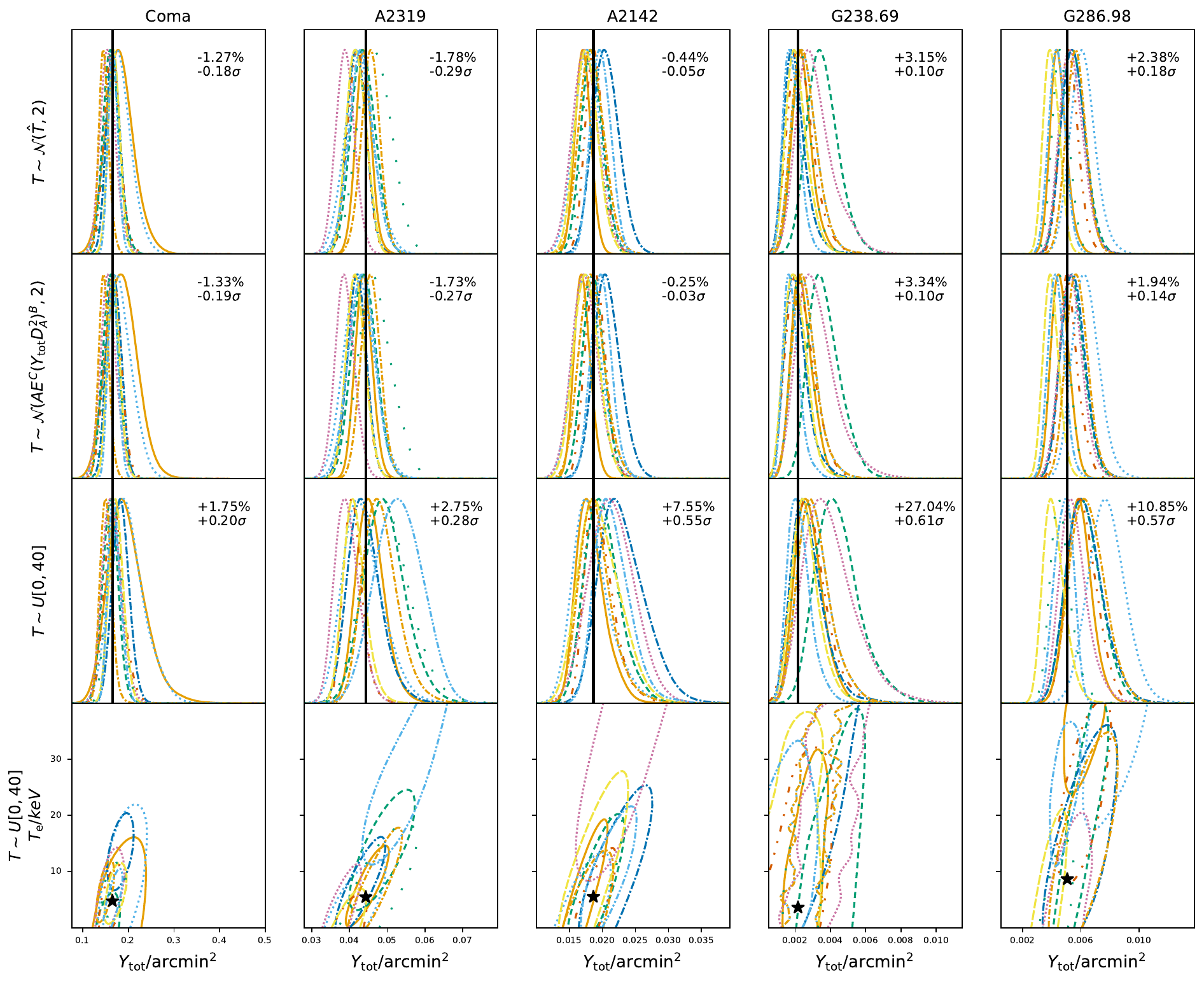}
    \caption{Posterior constraints on the cluster \Ytot\ and \Tsz\ model parameters derived from analysing the full relativistic simulations.  The bottom and second-to-bottom rows show two- and one-dimensional constraints when the cluster temperature is given a uniform prior, while the two top rows show the one-dimensional constraints when the cluster temperature is given an informative prior, either centred on the true temperature or on a scaling-relation-derived value (see text for more details).  Black vertical lines show true \Ytot\ values; other colours, markers and text are as in Figure~\ref{Fi:sim_posteriors}.}
    \label{Fi:sim_posteriors2}
\end{figure*}

In the higher SNR cases, the \Ytot\ measurements are unbiased when the uniform prior on \Tsz\ is used.  However, for most clusters, using the uniform prior introduces a bias in the \textit{opposite} direction (\Ytot\ is over-estimated).  This can be explained by the posterior distributions for \Ytot\ and \Tsz\ shown in Figure~\ref{Fi:sim_posteriors2}.  There is a strong degeneracy between these two parameters since the effect of the relativistic correction over the \emph{Planck} band is mostly to diminish the signal; there is also a small change in spectral slope and the position of the signal null but these effects are small given the \emph{Planck} noise levels and the change is not enough to break the degeneracy between \Ytot\ and \Tsz.  Marginalizing over the large range of \Tsz\ therefore induces a positive bias in the \Ytot\ marginal constraint.  In contrast, the two different informative priors on \Tsz\ produce unbiased \Ytot\ constraints at the $<3.5$\% and $<0.5\sigma$ level in all cases.

\begin{figure*}
\includegraphics[width=\linewidth]{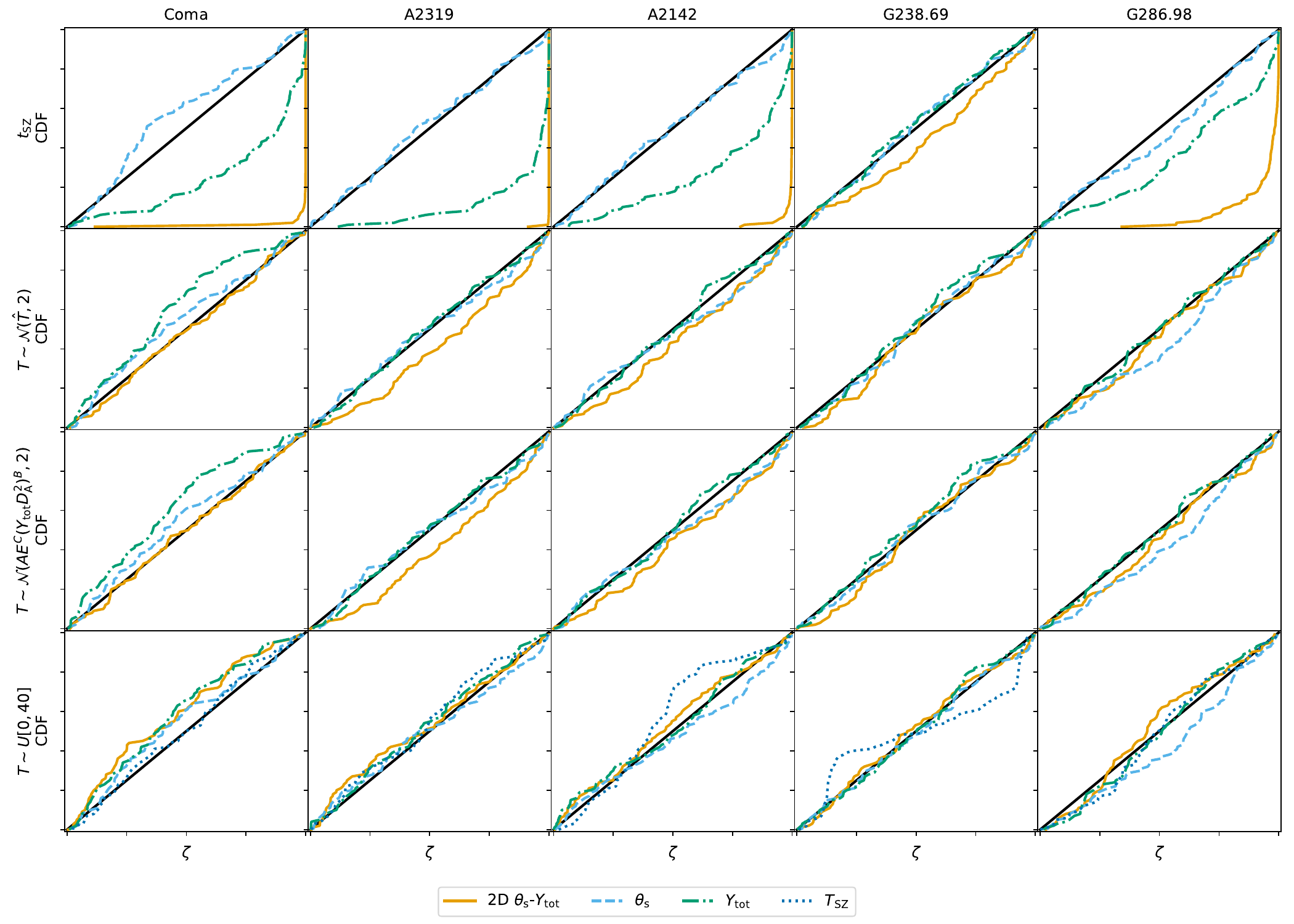}
    \caption{Posterior validation plots for the simulated clusters, assuming the non-relativistic approximation (top row), and an isothermal rSZ model with priors given in the axis labels (other rows).  All axis ranges are from $0$ to $1$.  The more closely the coloured posterior validation curves follow the black solid line, the more accurately the posterior describes the uncertainty on the parameter(s) of interest.  \Tsz\ validation curves are only shown for the uniform prior case, since with the informative priors the posterior is driven by the prior.}
    \label{Fi:sim_post_val}
\end{figure*}

Since there is a degeneracy between \thetas\ and \Ytot\ parameter constraints in the context of the GNFW model, external information is often used to break the degeneracy and obtain tighter constraints on \Ytot\ (and derived parameters such as $Y_{500}$).  For example, an X-ray measurement of $\theta_{500}$ can be used as an external prior on \thetas\ (in combination with an estimate of $c_{500}$), or a scaling relation between $\theta_{500}$ and $Y_{500}$ (via the relationship of both with $M_{500}$) can be used to `slice' the 2D \thetas-\Ytot\ posterior (again requiring an estimate of $c_{500}$, plus an estimate of the scaling relation parameters).  For this reason, it is important to ensure the accuracy of the 2D \thetas-\Ytot\ posterior as well as the 1D marginalized posteriors on the individual parameters.  We show posterior validation plots (see Section~\ref{S:posterior_validation}) for the cluster parameters in Figure~\ref{Fi:sim_post_val}.  It can be seen that only \thetas\ is measured correctly when the non-relativistic approximation is used.  When the relativistic spectrum is used, for all three prior options the 2D and 1D posteriors on \thetas\ and \Ytot\ are measured correctly by this metric.  It seems somewhat counterintuitive that the low-significance average downward bias in \Ytot\ in the tSZ case has a large effect on the posterior validation curves, whereas the (similarly) low-significance average upward bias in \Ytot\ in the rSZ case with uniform prior does not.  This is because including the marginalization over temperature enlarges the posteriors sufficiently to `hide' the bias.

The posterior validation curves for \Tsz\ in the uniform-prior case show that it is measured correctly (although weakly) in the case of the high-SNR clusters; going to lower-SNR it is not constrained so deviates from the expected behaviour since the posteriors encompass almost all values of \Tsz.  In the case of the informative priors, we do not expect the temperature posteriors to be measured correctly (they are driven by the priors), so we do not show the validation curve for temperature for these cases.

\subsubsection{Isothermality assumption}

\begin{figure}
\includegraphics[width=\linewidth]{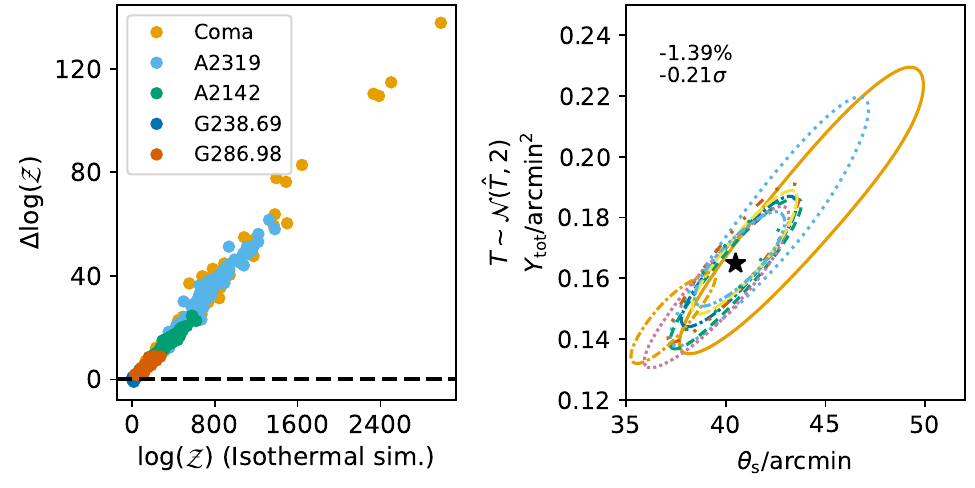}
    \caption{Left: comparison between log-evidence values for rSZ simulations \emph{created} using an isothermal cluster model and a radially varying temperature profile, but both \emph{analysed} using an isothermal model.  The $x$-axis shows the log-evidence, indicating the detection significance of the cluster, when the cluster is simulated and analysed using an isothermal model.  The $y$-axis shows the \emph{difference} in log-evidence, for each realization in each simulation set, between the $x$-axis value and the log-evidence obtained when the cluster is \emph{simulated} with a radially varying temperature profile but \emph{analysed} with an isothermal model; ie the log-evidence values are higher by up to $\approx$\,100 when the isothermal model is used both to create and analyse the simulations.  In both cases, the prior on temperature is centred on the true, pressure-weighted average value.  Right: posteriors for 10 simulation realisations for the Coma-like cluster, where the isothermal model has been used both to create and analyse the simulations.  The realizations are the same as those shown in Figure~\ref{Fi:sim_posteriors}.}
    \label{Fi:isothermal_evs}
\end{figure}

We tested the impact of the isothermality assumption by making two different versions of the relativistic simulations for each cluster: one with a radially varying temperature profile, and another with a single average temperature.  We analysed each simulation set in the same way and compared the results, finding that the evidence values were systematically higher for the isothermal simulations, particularly for the higher-evidence detections.  This is shown on the left-hand side of Figure~\ref{Fi:isothermal_evs}.  This indicates that the isothermal simulations are better-fitted by the isothermal model, even given the limitations of the \emph{Planck} data.

However, the overall results for \Ytot\ and \thetas\ did not change significantly, as illustrated in the right-hand plot in Figure~\ref{Fi:isothermal_evs}.  This shows the same ten noise realizations as Figure~\ref{Fi:sim_posteriors} for the Coma-like cluster, analysed with the Gaussian prior on temperature centred on the true value.  The posteriors are visually slightly more centred on the true values of \Ytot\ and \thetas, however overall the bias in \Ytot\ with respect to the true value is not significantly different, even though the log-evidence difference between simulation types is very large for this cluster.  We therefore conclude that although a non-isothermal model may be a better fit to the data, it will not improve the integrated Compton-$y$ estimate and we do not explore this idea further in this paper.

\subsubsection{Temperature systematics}\label{S:temp_systematics}

A related question is, what level of systematic error is introduced when the temperature estimate used for the informative prior is biased in some way?  For example, if an X-ray temperature estimate is used, it is likely to be a cylindrically-averaged estimate within $\sim r_{500}$, whereas our \Tsz\ is averaged within $\approx 5r_{500}$.  We used the \citet{2006ApJ...640..710V} method for estimating X-ray temperatures in conjunction with our physical model for the electron density and temperature to predict the ratio between the X-ray and SZ temperatures for a range of masses and redshifts, both for \emph{Chandra} and XMM-Newton.  We found that the X-ray temperature within $r_{500}$ is biased upward by a maximum of $\approx$\,40\% compared to \Tsz\ if the core is excised, for a \emph{Chandra} measurement.  This difference is mainly due to the decrease in the temperature profile outside of $r_{500}$.  We tested the effect of this by rerunning the analysis on our set of simulated clusters, with the Gaussian prior on temperature centred on $1.4 \times \Tszm$.  The effect of the biased temperature is most pronounced for the hottest cluster, G286.98, where a 40\% increase in temperature results in a 5\% increase in \Ytot\ on average.  In contrast, for the A2319-like cluster, a 40\% increase in temperature results in only a 3\% average increase in \Ytot.

On the other hand, the best currently available \Tsz\ scaling relations are those of \citet{2020MNRAS.493.3274L} and \citet{2022MNRAS.517.5303L}, which give relationships between mass and pressure-weighted average temperature within $r_{500}$ and $r_{200}$ respectively based on numerical simulations.  We compared the spherical averages to our total pressure-weighted temperatures using our model, finding that these temperatures are biased high by a maximum of $\approx$\,30\% ($T_{500}$) and $\approx$\,20\% ($T_{200}$).  We again tested these biases on our simulation set and found that for G286.98, a 30\% (20\%) increase in temperature results in a 4\% (2.5\%) increase in \Ytot\ on average.  The $M_{200}/T_{200}$ scaling relation should therefore be the safest to use in lieu of any real temperature measurements sensitive to the outskirts of the cluster; we note also that our simple model is not necessarily a good representation of the real physics in the outskirts of clusters, where complexities occur due to accretion etc.  We will return to this issue in Section~\ref{S:relativistic_SR} where we use X-ray and simulation-based temperatures for our scaling relation calibration.

\subsubsection{Correlated dust emission}\label{S:correlated_dust}

A dust component has been detected covering a similar spatial extent to the ICM, thought to originate either from the galaxies in the clusters or from a diffuse intracluster dust component (e.g.\ \citealt{2008A&A...490..547G}, \citealt{2016A&A...594A..23P}, \citealt{2016A&A...596A.104P}, \citealt{2018MNRAS.476.3360E}).  We tested the effect of this by inserting a Gaussian dust component with $\sigma = \theta_{500}$ on top of the cluster signal in each simulated frequency map.  The spectrum and amplitude of the dust component follow the average constraint from \citet{2018MNRAS.476.3360E}.  Then, we repeated the same analysis as in the no-dust simulations.  The results show negligible change to the recovered parameter values and only a small decrease in evidence overall when the dust component is present, even in the case of the non-informative prior on temperature.  This indicates that the \textsc{PwS} methodology is able to robustly separate out the dust component and include it in the generalized noise estimate based on its different spectrum, even when the SZ spectrum is changing due to temperature.

\subsubsection{Summary of simulation analysis}

In summary, based on the analysis of the simulations we conclude that neglecting the rSZ corrections results in an integrated Compton-$y$ estimate that is biased down with respect to the true value by an amount ranging from $\approx$\,5\% to 15\% depending on the mass (and therefore temperature) of the cluster.  While the bias is only significant at $\approx$\,3$\sigma$ for the highest-SNR clusters, when averaging over a sample of clusters to constrain a scaling relation there will be a global mass-dependent bias.

We can weakly constrain \Tsz\ for the highest-SNR clusters, however for the bulk of the sample an informative prior is required to correct the rSZ bias and produce a correct posterior constraint on \thetas\ and \Ytot.  The isothermal assumption is accurate enough for this analysis; systematic differences in external temperature estimates may be an issue and will be considered further when analysing real data; and correlated dust emission produces a negligible effect.

\section{Pressure profile shapes}\label{S:profile_shapes}

A secondary effect we wish to explore is the effect of varying pressure profile shapes on the integrated Compton-$y$ estimates from \emph{Planck}.  We used the X-ray constraints from the L20 sample to investigate this issue.

\begin{figure}
\includegraphics[width=\linewidth]{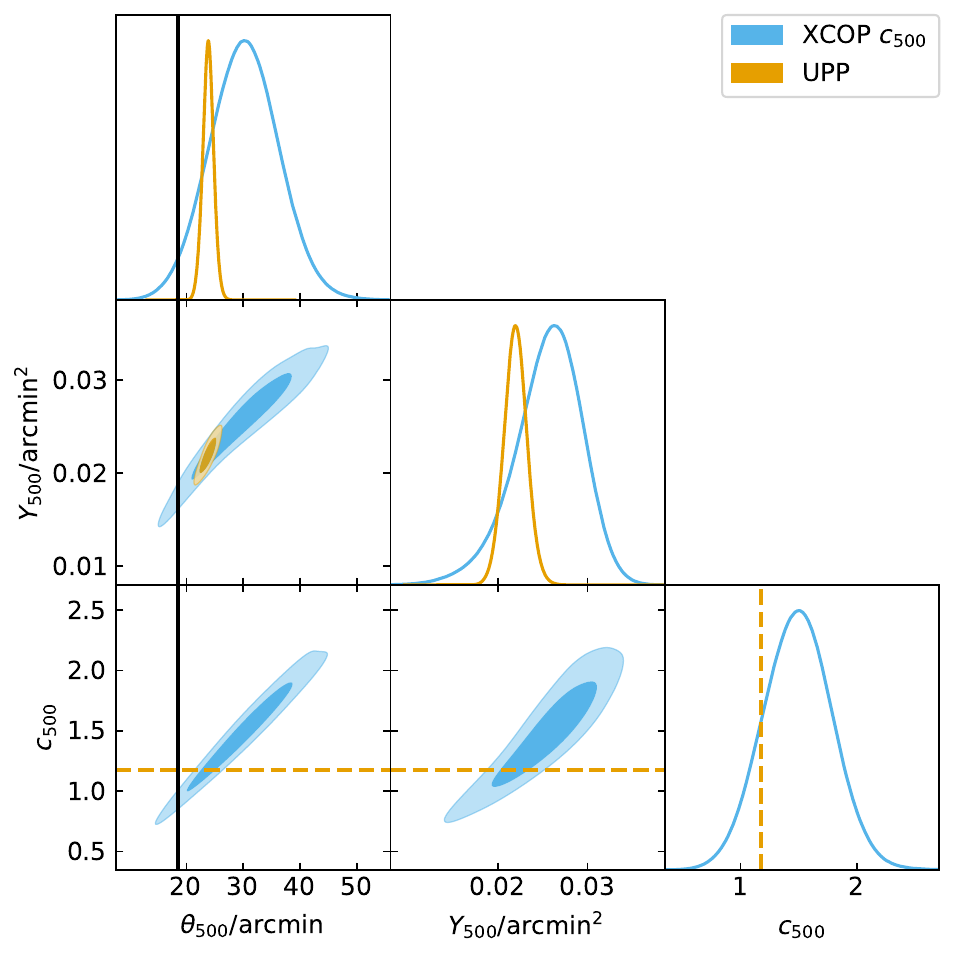}
    \caption{Comparison between posterior constraints on $\theta_{500}$ and $Y_{500}$ for the high-significance cluster Abell 3266 using the (fixed) UPP value of $c_{500}$ compared to the XCOP value with error.  The black solid lines show the X-ray measured value of $\theta_{500}$ from L20 and the orange dashed line shows the UPP value of $c_{500}$.}
    \label{Fi:A3266_c500}
\end{figure}

Firstly, using the high-significance cluster Abell 3266 (PSZ2 G272.08-40.16) as an example, we illustrate the effect of varying $c_{500}$ in Figure~\ref{Fi:A3266_c500}.  We fit the real \emph{Planck} data using the observational GNFW model with the uninformative priors on $x_0$, $y_0$, \thetas\ and \Ytot\ given in Table~\ref{T:priors}.  We firstly fixed all shape parameters (including $c_{500}$) to the UPP values; then left \cGNFW, \aGNFW, \bGNFW\ fixed to the UPP values but gave $c_{500}$ a Gaussian prior based on the XCOP value and error.  When $c_{500}$ is fixed to its UPP value, the SZ constraints are significantly inconsistent with the X-ray value for $\theta_{500}$, while the uncertainty allowed by the XCOP prior brings them within (better) consistency.  We note that there is no information on $c_{500}$ in the \emph{Planck} data in the context of the observational model, so $c_{500}$ is just an externally-imposed factor which translates from \thetas\ to $\theta_{500}$ (and from \Ytot\ to $Y_{500}$).

\begin{figure}
\includegraphics[width=\linewidth]{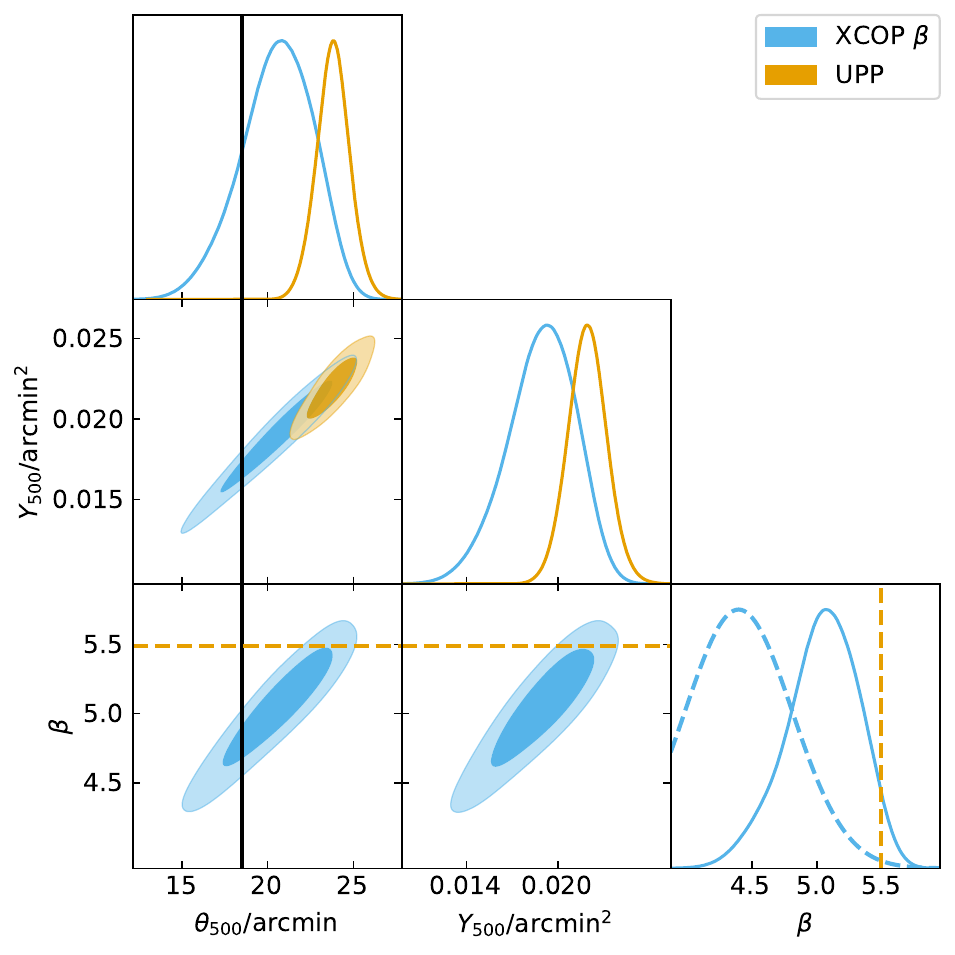}
    \caption{Comparison between posterior constraints on $\theta_{500}$ and $Y_{500}$ for the high-significance cluster Abell 3266 using the (fixed) UPP value of $\beta$ compared to the XCOP value with error.  The black lines show the X-ray measured value of $\theta_{500}$ from L20 and the orange dashed line shows the UPP value of $\beta$.  The blue dashed line shows the Gaussian XCOP prior on $\beta$.}
    \label{Fi:A3266_beta}
\end{figure}

Similarly, we illustrate the effect of changing the prior on $\beta$ from the fixed UPP value to a Gaussian prior based on the XCOP estimate and error in Figure~\ref{Fi:A3266_beta}.  The change in $\beta$ also brings the SZ $\theta_{500}$ estimate into much better agreement with the X-ray value.  In this case, there is information in the \emph{Planck} data on the $\beta$ value as shown by the movement away from the prior.

\begin{figure}
\includegraphics[width=\linewidth]{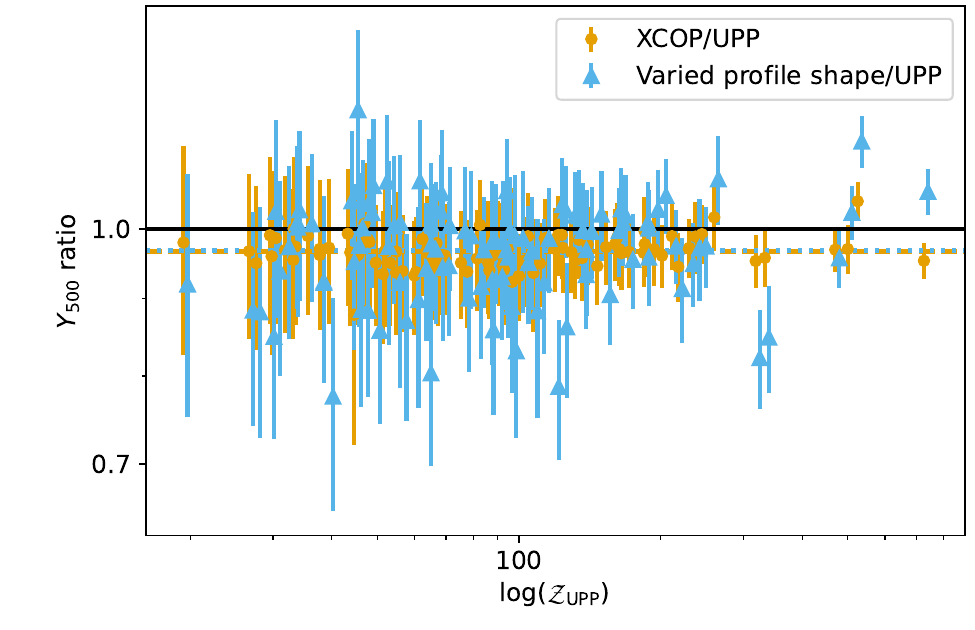}
    \caption{Comparison between $Y_{500}$ measurements obtained using a fixed UPP profile; a fixed XCOP profile; and varying pressure profile parameters with a prior on $\theta_{500}$ based on the L20 X-ray mass.  The $x$-axis shows log-evidence values from the UPP analysis; the $y$-axis shows the ratio between $Y_{500}$ measurements using the XCOP profile and UPP profile with yellow circles, and the corresponding ratio when varying the pressure profile with blue triangles.  Points have been slightly displaced horizontally for clarity.  The overall mean ratios are shown with yellow dashed and blue dotted horizontal lines and are barely distinguishable at $\approx$\,4\%.}
    \label{Fi:profile_comparison}
\end{figure}

Since there is clearly information on the profile shape in the \emph{Planck} data, we also tried a uniform prior on each of \aGNFW, \bGNFW, $c_{500}$, leaving \cGNFW\ fixed to the XCOP mean since the relatively low-resolution \emph{Planck} data is less sensitive to this parameter.  The uniform prior limits are shown in Table~\ref{T:priors}.  There are too many degeneracies present in the model to fit all parameters simultaneously, however switching to the physical overdensity model and using an X-ray prior on $M_{500}$ (which translates to a prior on $\theta_{500}$; see Section~\ref{S:SR_model}) the degeneracies are reduced sufficiently to achieve a good $Y_{500}$ measurement.  Fig~\ref{Fi:profile_comparison} shows a comparison between $Y_{500}$ as measured from the \emph{Planck} data using (i) the observational GNFW model with a fixed UPP profile; (ii) the observational GNFW model with a fixed XCOP profile; and (iii) the physical overdensity model, varying profile shape parameters with an L20 prior on $M_{500}$.  In the case of both (ii) and (iii), there is an overall shift down in $Y_{500}$ by $\approx$\,4\% when compared to results with the fixed UPP profile.  While this is a small change and not significant on the individual cluster level, similarly to the rSZ effect it is a source of systematic error that should be taken into consideration in the quest for precision cluster measurements.

\begin{figure}
\includegraphics[width=\linewidth]{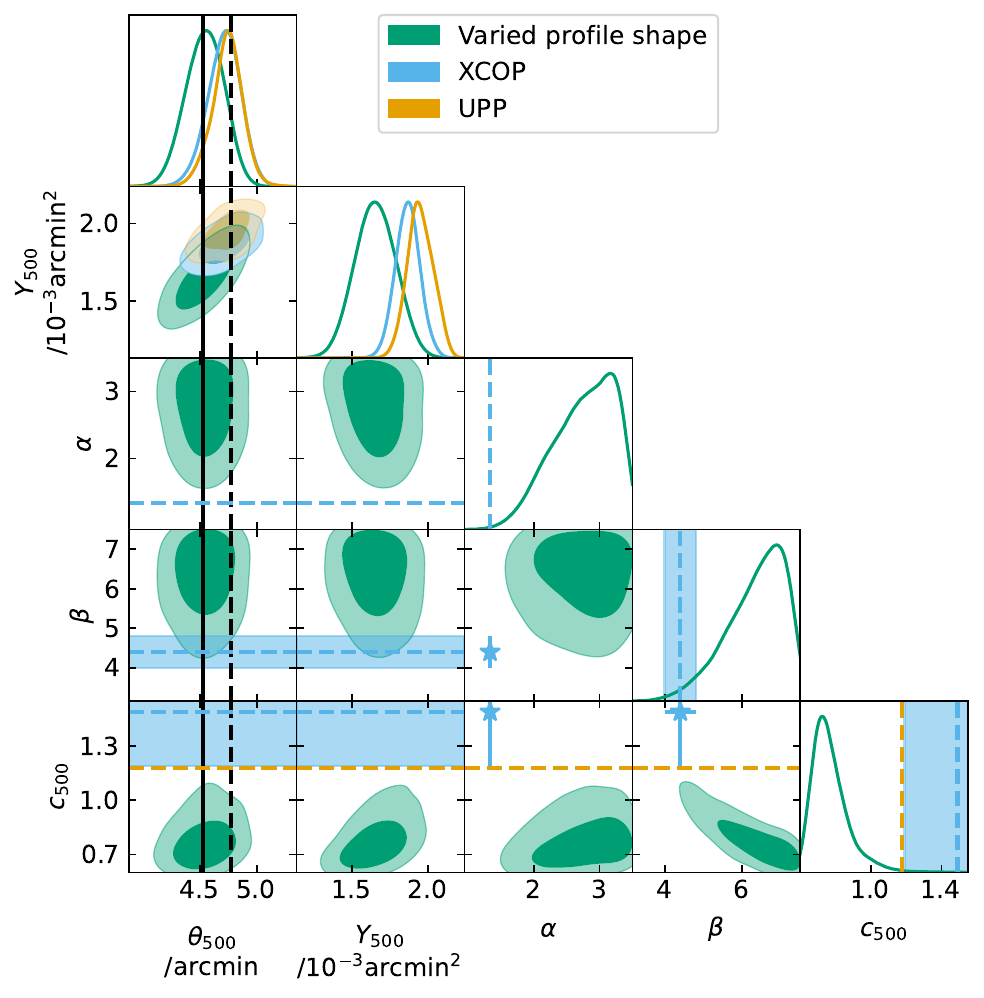}
    \caption{Comparison between posterior constraints on $Y_{500}$ and the GNFW profile parameters for the high-significance cluster Abell~S~520, using an X-ray prior on $\theta_{500}$.  XCOP values and errors are illustrated with dashed blue lines, bands and stars, and UPP values with dashed orange lines and stars.  The X-ray constraint on $\theta_{500}$ is shown with a black solid line (L20) and dashed line \citep{2011A&A...534A.109P}.  The strong departure from the average profile parameters appears, in this case, to be adjusting the profile shape to accommodate the low $\theta_{500}$ estimate from L20.}
    \label{Fi:AS520_uniform}
\end{figure}

For most clusters in the L20 sample, the constraints on the profile shape parameters are consistent with the XCOP profile and the corresponding $Y_{500}$ measurements are also consistent (with larger error bars when the profile shape is varied).  However, in a few high-significance cases, the results do change.  This is illustrated for the most extreme case (PSZ2~G262.27-35.38 or Abell~S~520) in Figure~\ref{Fi:AS520_uniform}, where the profile shape parameter constraints are strongly discrepant with the XCOP (and UPP) values.  In this case, we interpret this to mean that in fact the mass estimate from L20, and therefore the prior constraint on $\theta_{500}$, is too low; by adjusting $c_{500}$ and \bGNFW, the profile can be adjusted to fit the \emph{Planck} data better while still matching the $\theta_{500}$ prior.  This is the desired behaviour for calibrating a scaling relation; in contrast, when the profile shape is fixed the $\theta_{500}$ posterior is forced to depart from the X-ray prior.  This interpretation may be supported by other literature values for the mass (e.g.\ $M_{X,500}=7.7 \times 10^{14} M_{\odot}$ from \citet{2011A&A...534A.109P}, compared to $6.59 \times 10^{14} M_{\odot}$ from L20 (although note that the \citealt{2011A&A...534A.109P} mass is derived from an $L_X-M$ scaling relation and no uncertainties are given).  In the case of other, high-significance clusters such as Abell 3266, the opposite is true: the $\theta_{500}$ value does not change, but the profile shape parameter constraints fall in slightly different regions to the XCOP or UPP values and therefore the recovered $Y_{500}$ value does change significantly.

We have verified with simulations based on the L20 clusters that $Y_{500}$ is recovered correctly when we allow the profile shape parameters to vary with the uniform priors.  We note that the \textsc{PwS} initial background subtraction step requires a fixed profile shape, which we set to XCOP.  In the cases we tested, we generated the simulations with a UPP profile to verify that the mismatch in the profile in the background subtraction step did not affect the final constraints; since there are large degeneracies in all the parameters, it is generally possible to find a sufficiently good fit to the data even with the wrong profile, although the \thetas\ and \Ytot\ estimates (in the background subtraction step only) will be incorrect.

\subsection{Effect of \cGNFW}

\begin{figure}
\includegraphics[width=\linewidth]{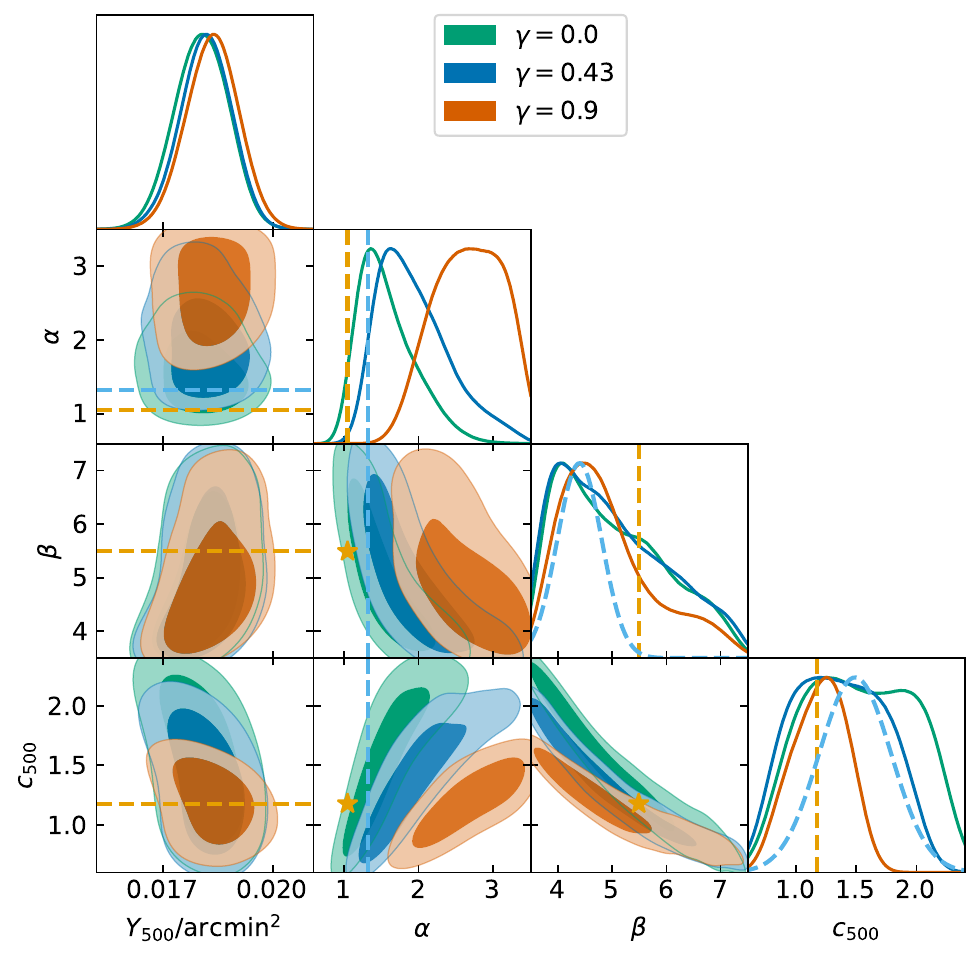}
    \caption{Comparison between posterior constraints on $Y_{500}$ and the GNFW profile parameters for the high-significance cluster Abell 3266, using an X-ray prior on $\theta_{500}$, and three different fixed values of \cGNFW\ as indicated in the legend.  The change in \cGNFW\ has an insignificant impact on the $Y_{500}$ constraint despite the changes in the profile shape parameter posteriors.  XCOP values and priors are illustrated with dashed blue lines and UPP values with dashed orange lines and stars.}
    \label{Fi:A3266_test_gamma}
\end{figure}

\emph{Planck} data is generally thought to be of too low resolution to be impacted by the \cGNFW\ parameter which describes the innermost part of the cluster pressure profile, and we do not vary it in our analysis.  However, for clusters such as Abell 3266 with $\theta_{500}\approx$18\,arcmin, the cluster is large compared to the $\approx$5\,arcmin resolution of the higher-frequency \emph{Planck} channels.  We tested whether \cGNFW\ has a significant impact on the $Y_{500}$ constraints by performing separate analyses with the uniform priors on \aGNFW, \bGNFW\ and $c_{500}$ and the X-ray-based prior on $\theta_{500}$, fixing \cGNFW\ to extreme values of $0$ and $0.9$, $\approx\,4\sigma$ from the XCOP mean.  The resulting posteriors are shown in Figure~\ref{Fi:A3266_test_gamma}.  It is clear that the $Y_{500}$ constraint is not significantly impacted (it shifts by $<1$\%), although there is information in the data on the \cGNFW\ parameter with shape parameter constraints shifting and the $\cGNFWm=0.0$ model being preferred with respect to the $\cGNFWm=(0.43,0.9)$ models with log-evidence differences of $\Delta \mathcal{Z} = (0.5, 5.1)$.  Since we are mostly concerned with integrated $Y$ we do not explore this any further and leave \cGNFW\ fixed to the XCOP mean of $0.43$ for further analysis.

\subsection{Choice of parameter prior limits}\label{S:GNFW_priors}

It should be noted that Figures~\ref{Fi:AS520_uniform} and \ref{Fi:A3266_test_gamma} show that some of the posteriors for the GNFW shape parameters hit the prior limits.  This is due to the severe parameter degeneracies inherent in the GNFW model, particularly when only a limited range of angular scales are constrained in the profile.  Experimentation with widening the prior limits beyond the ranges given in Table~\ref{T:priors} has shown that the $N$-dimensional posterior shapes become complex and difficult to interpret.  The limits used here restrict the parameter values to ranges which are sensible based on our prior knowledge of the average pressure profile shape and its dispersion (eg from studies such as \citealt{2023ApJ...944..221S} and \citealt{2019A&A...621A..41G}) while not including the parameter ranges where the posterior shape becomes complex.  As noted above, we have verified with a simulation set based on the L20 cluster sample that $Y_{500}$ is recovered correctly when we allow the profile shape parameters to vary with uniform priors with these ranges, even though the posterior constraints hit the prior boundaries.  These simulations were created with UPP profiles, analyzed with XCOP profiles in the background estimation step, and then analyzed with the uniform priors for parameter estimation.  In this test, recovered $Y_{500}$ mean values were unbiased to within $<3$\%, and the posterior validation test showed the errors on $Y_{500}$ were also accurate.  The $\theta_{500}$ posteriors were entirely driven by the external prior and so mean $\theta_{500}$ is also recovered correctly.  In contrast, when analyzing these cluster simulations with an (incorrect) fixed XCOP profile, mean $Y_{500}$ values were biased down by $\approx$\,10\%.  We can therefore be confident that the restriction on the prior ranges is not producing a bias in $Y_{500}$, for reasonable true pressure profile shapes.

\subsection{Summary of pressure profile investigation}

Based on the investigation here, we conclude that accurate constraints on integrated $Y$ from \emph{Planck} data require \textit{either} accurate external constraints on the profile shape, \textit{or} external information on a characteristic scale such as $\theta_{500}$.  This could be from an external measurement or via a scaling relation which imposes an expected relationship between $\theta_{500}$ and $Y_{500}$.  For the remaining analysis in this paper, we will allow the pressure profile parameters to vary using the uniform priors shown in Table~\ref{T:priors} since we have external measurements of $\theta_{500}$ from L20.

\section{Mass scaling relation calibration}\label{S:scaling_relations}

We follow \emph{Planck} analysis in calibrating the relationship between $M_{X,500}$ as measured from X-ray data assuming hydrostatic equilibrium, and $Y_{500}$ as measured from \emph{Planck} data.  We fit a scaling relation of the same form used in the \emph{Planck} collaboration analysis:

\begin{equation}\label{eq:scaling_relation}
E^{-B}(z) \left [ \frac{D_\mathrm{A}^2 Y_{500}}{10^{-4} \mathrm{Mpc}^2} \right ] = Y_{*} \left [ \frac{h}{0.7} \right ]^{-2+A} \left [ \frac{M_{X,500}}{6 \times 10^{14} \mathrm{M}_{\odot}} \right ]^A,
\end{equation}
where $M_{X,500}$ is the X-ray-derived hydrostatic equilibrium mass which may be biased compared to the true mass.  In the self-similar model, $B=2/3$ and $A=5/3$.

We focus first on the non-relativistic case, and validate our updated \emph{Planck} measurements against the results from \citet[hereafter P13]{2014A&A...571A..20P} to check for any global calibration offsets.  We then extend the sample to lower mass, using the hydrostatic masses measured by L20, and recalibrate the relationship.

For our SZ measurements, we carry out the analysis of the NPIPE \emph{Planck} data with the physical overdensity model, using an asymmetric Gaussian prior on $M_{X,500}$ derived from the L20 estimate and upper and lower errorbars.  As noted in Section~\ref{S:profile_shapes}, as long as the profile shape is allowed to vary this means that the posterior on $\theta_{500}$ does not depart from the prior, so we are measuring integrated $Y$ inside the radius defined by the X-ray mass, taking into account its uncertainty.  We use uniform priors on the GNFW parameters and a non-informative log-uniform prior on $Y_{500}$ as described in Table~\ref{T:priors}.

We use the LInear Regression in Astronomy (LIRA; \citealt{2016MNRAS.455.2149S}) package to perform the scaling relation fits.  This method can take into account errors (possibly correlated) in both measured parameters as well as intrinsic scatter in both.  

\subsection{Comparison with P13}

P13 used a subset of 71 high-SNR clusters from the PSZ1 catalogue which had good-quality \emph{XMM-Newton} observations.  The SZ signal was re-extracted from the 15.5 month \emph{Planck} survey data, centred on the position of the X-ray peak and within the $r_{500}$ radius determined from X-ray data.  A matched multi-filter method was used, assuming the UPP, and the resulting $Y_{500}$ were corrected for Malmquist bias.  The scaling relation was fit using the orthogonal BCES method \citep{1996ApJ...470..706A}, taking into account uncertainties in both variables and (a single value of) intrinsic scatter.

We first tested that we could recover consistent scaling relation parameter fits using LIRA and the original bias-corrected P13 data\footnote{Data obtained from \url{http://szcluster-db.ias.u-psud.fr}.}.  The fits are shown in Table~\ref{T:scaling_relation_fits}, along with the fitting results reported by P13.  The results are extremely consistent showing that any differences in scaling relation calibration are not due to the fitting methodology used.

\begin{figure}
\includegraphics[width=\linewidth]{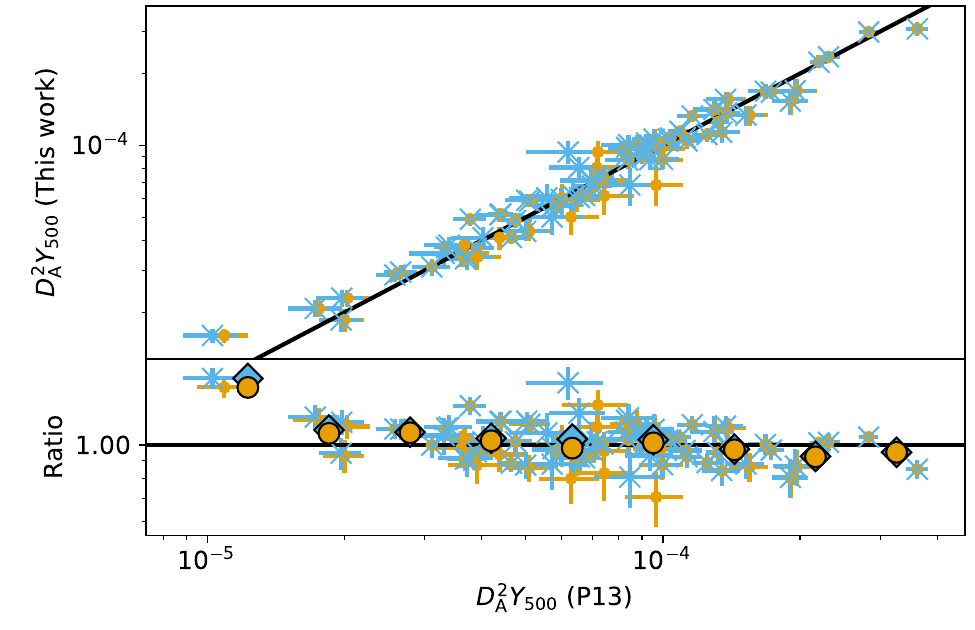}
    \caption{Comparison between our posterior constraints on $Y_{500}$ using P13 X-ray mass priors, and the $Y_{500}$ values used by P13.  The blue crosses (small yellow dots) show data points with (without) the Malmquist bias correction applied by P13. In the lower axis, the large blue diamonds (large yellow circles) show the corresponding average ratios in bins in $D_\mathrm{A}^2 Y_{500}$.}
    \label{Fi:P13_Y500_comp}
\end{figure}

We next compared the P13 $Y_{500}$ values to our updated values.  We discarded cluster PSZ1 G282.45+65.18 which was not found in the PSZ2 cosmological sample, leaving a comparison sample of 70.  The $Y_{500}$ comparison is shown in Figure~\ref{Fi:P13_Y500_comp}; the measurements are extremely consistent over most of the $Y_{500}$ range, showing that there is no global offset introduced by the change of \emph{Planck} dataset and/or fitting methodology.  Our updated values at small $D_\mathrm{A}^2 Y_{500}$ do appear to be systematically higher than the P13 values, although there are few clusters in this range.

\subsection{Updated calibration}

\begin{figure}
\includegraphics[width=\linewidth]{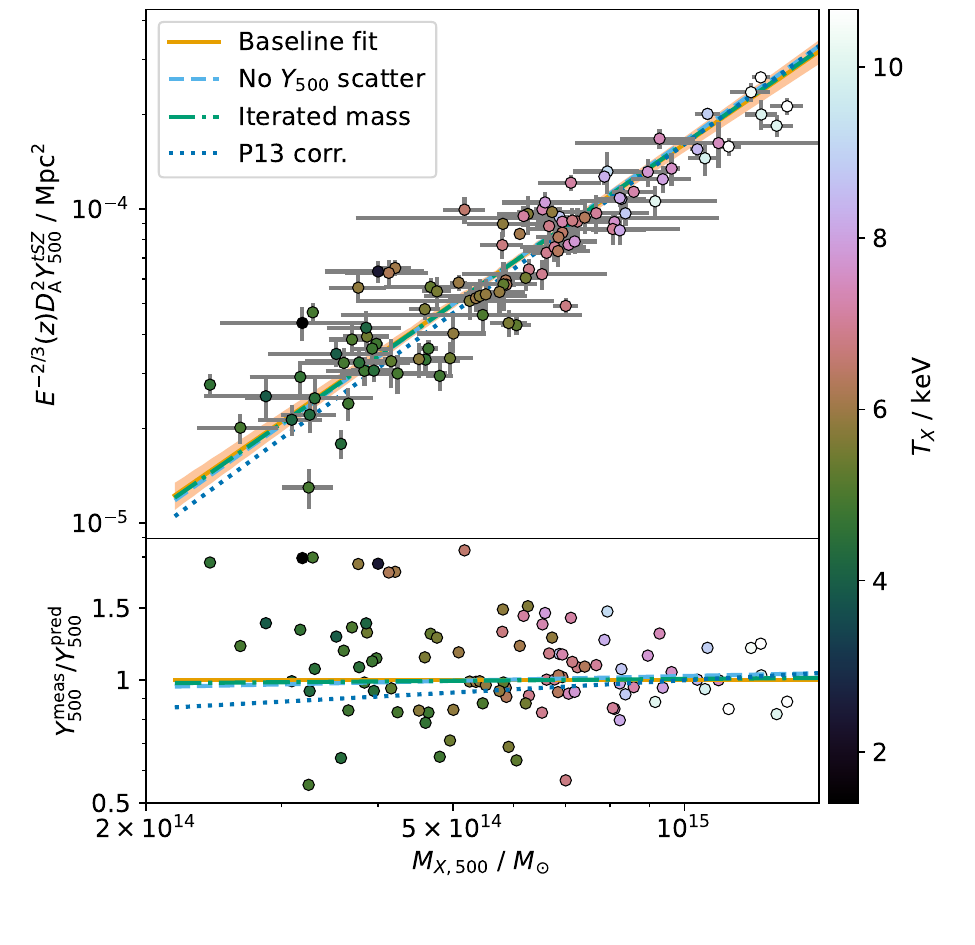}
    \caption{Baseline calibration of the mass-observable scaling relation assuming the non-relativistic SZ spectrum.  Our calibration (orange line, shaded area showing uncertainty) is consistent within $\approx$\,$1\sigma$ with the result from P13 (blue dotted line) over most of the mass range, deviating slightly at the low-mass end.  The dashed light blue line shows the fit when only allowing for scatter in the mass observable, and the green dot-dashed line shows the fit when a bootstrapped mass consistent with the scaling relation is used to define $\theta_{500}$ for the SZ signal extraction.}
    \label{Fi:SR_fits}
\end{figure}

We use the hydrostatic masses measured by L20 for our subsample of 101 clusters, as described in Section~\ref{S:Xray_data}.  We use LIRA to perform the fit, and report the results of all scaling relation fits in Table~\ref{T:scaling_relation_fits}.  As a baseline fit, we fix $B$ to its self-similar value and allow intrinsic scatter in both parameters.  Figure~\ref{Fi:SR_fits} shows this baseline scaling relation fit in comparison to the \citet{2014A&A...571A..20P} result (which did not change in subsequent \emph{Planck} analysis); the fits are extremely consistent at the high-mass end, and only deviate by slightly more than $1\sigma$ at the low-mass end.  There are several differences between our analysis methodologies: (i) we use a larger sample of clusters (with some overlap); (ii) we use hydrostatic equilibrium X-ray masses instead of masses derived using a $Y_{X}-M_{X,500}$ scaling relation; (iii) we use \textsc{PwS} rather than a matched filtering method to constrain $Y_{500}$; (iv) we allow for variation in the cluster pressure profile shape; (v) we use the updated NPIPE reduction of the \emph{Planck} data with greater depth.  This consistency therefore serves to highlight the robustness of this particular mass-observable scaling relation.  Table~\ref{T:scaling_relation_fits} also shows that our fit is extremely consistent with the \emph{Planck} result, with both the slope and normalisation consistent at $\approx$\,$1\sigma$.

\begin{table*}
\centering
\caption{Scaling relation fitting results, using symbols as defined in equation~\ref{eq:scaling_relation}.  Parameters with errorbars of $0.0$ are fixed to the given value.  `P13 corr.' refers to the Malmquist-bias-corrected scaling relation result from \citet{2014A&A...571A..20P} (using BCES) and the LIRA fit is our own fit to the P13 data.  `tSZ' and `rSZ' refer to results from this work for the non-relativistic and relativistic SZ spectra respectively.}
\label{T:scaling_relation_fits}
\begin{tabular}{lcccccccccc}
\hline
Type & $A$ & $\Delta A$ & $B$ & $\Delta B$ & $\log Y_{*}$ & $\Delta \log Y_{*}$ & $\sigma_{\log M}$ & $\Delta \sigma_{\log M}$ & $\sigma_{\log Y}$ & $\Delta \sigma_{\log Y}$ \\
\hline
P13 corr. (BCES) & 1.790 & 0.060 & 0.667 & 0.000 & -0.190 & 0.010 & - & - & $<0.074$ & -\\
P13 corr. (LIRA) & 1.783 & 0.067 & 0.667 & 0.000 & -0.182 & 0.011 & 0.036 & 0.006 & <0.024 & -\\
tSZ (baseline) & 1.682 & 0.087 & 0.667 & 0.000 & -0.168 & 0.013 & 0.054 & 0.008 & <0.046 & -\\
\textbf{tSZ, no $Y_{500}$ scatter} & \textbf{1.728} & \textbf{0.080} & \textbf{0.667} & \textbf{0.000} & \textbf{-0.167} & \textbf{0.013} & \textbf{0.060} & \textbf{0.006} & \textbf{0.000} & \textbf{0.000}\\
tSZ, iterated mass & 1.705 & 0.119 & 0.667 & 0.000 & -0.168 & 0.015 & >0.057 & - & <0.069 & -\\
tSZ, SNR$>10$ & 1.633 & 0.091 & 0.667 & 0.000 & -0.161 & 0.013 & 0.057 & 0.008 & <0.038 & -\\
tSZ, completeness$>0.9$ & 1.663 & 0.108 & 0.667 & 0.000 & -0.164 & 0.015 & 0.063 & 0.008 & <0.042 & -\\
tSZ, free $B$ & 1.730 & 0.127 & 0.331 & 0.639 & -0.139 & 0.013 & 0.055 & 0.008 & <0.041 & -\\
\textbf{rSZ, $M_{200}-T_y$, no $Y_{500}$ scatter} & \textbf{1.765} & \textbf{0.081} & \textbf{0.667} & \textbf{0.000} & \textbf{-0.130} & \textbf{0.013} & \textbf{0.058} & \textbf{0.006} & \textbf{0.000} & \textbf{0.000}\\
rSZ, $M_{500}-T_y$, no $Y_{500}$ scatter & 1.769 & 0.080 & 0.667 & 0.000 & -0.127 & 0.013 & 0.058 & 0.006 & 0.000 & 0.000\\
rSZ, $T_X$, no $Y_{500}$ scatter & 1.773 & 0.081 & 0.667 & 0.000 & -0.125 & 0.013 & 0.059 & 0.006 & 0.000 & 0.000\\
\hline
\end{tabular}

\end{table*}

From this baseline fit, we experiment with variations in the fitting methodology and use a Monte-Carlo simulation in order to determine which of the variations that make a significant difference we should use to fit the data in the most robust way.

\subsubsection{Intrinsic scatter constraints}\label{S:intrinsic_scatter}

\begin{figure}
\includegraphics[width=\linewidth]{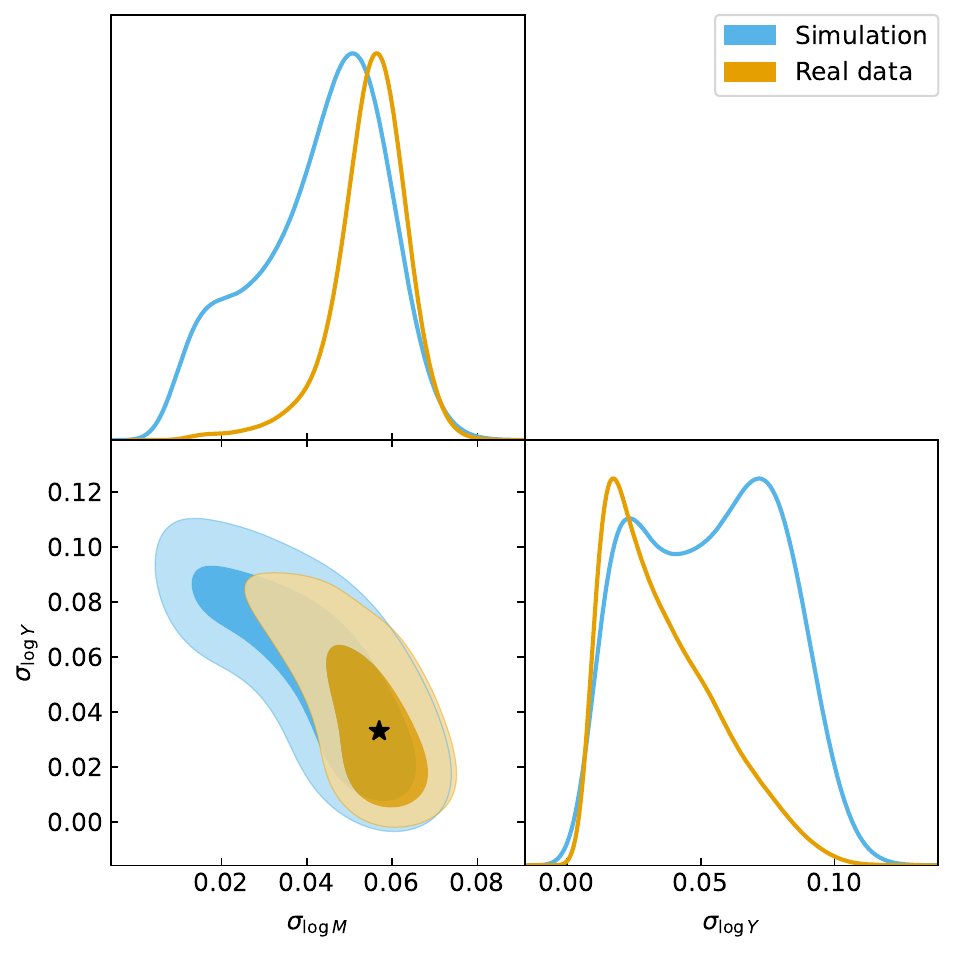}
    \caption{Posterior distributions of the intrinsic scatter parameters returned by LIRA, for our real sample data and a simulation, created as described in Section~\ref{S:selection_function}.  The simulation was created using the mean posterior values of the scatter parameters from the real data fit, marked with the black star.}
    \label{Fi:intrinsic_scatter}
\end{figure}

In our baseline fit, we used LIRA to estimate the intrinsic scatter in both coordinates, assuming a log-normal distribution for the scatter.  However, we find from inspecting the posterior distributions that the two scatter estimates are correlated, as shown in Figure~\ref{Fi:intrinsic_scatter}, where we show the LIRA posterior distributions for the intrinsic scatter parameters both for the real data, and for a simulation (see Section~\ref{S:selection_function}).  The simulation was created using input values of the intrinsic scatter based on the posterior means from the real data fit, and similar measurement errors to the real data.  It is clear that a simple 1D mean is inadequate to describe the posterior in the simulation case.  In the real data case, the posterior converges toward the lower edge of the inverse $\chi^2$ prior in $\sigma_{\log{Y}}$ and gives an upper 68\% confidence limit of 0.05.  This corresponds to a percentage error of $\approx$10\% which is consistent with the mean uncertainty in $Y_{500}$ of around 10\%; it is clearly difficult to constrain an intrinsic scatter level of less than the measurement error.

\subsubsection{Intrinsic scatter in mass only}

Since the scatter in $Y_{500}$ may be below the detection threshold, we considered only fitting for intrinsic scatter in $M_{X,500}$.  This results in the fit parameters shown in Table~\ref{T:scaling_relation_fits}, with an increase in the slope parameter $A$ of $\approx$\,0.5$\sigma$, a negligible change to the normalization $Y_*$ and a slight increase in the intrinsic mass scatter of $\approx$\,0.7$\sigma$.  We will show in Section~\ref{S:selection_function} that this fit is slightly more robust than the original fit for the situation where scatter in $Y_{500}$ is at or below the detection threshold, so we adopt this as our fiducial fit even though it is not significantly different to the baseline fit.

\subsubsection{Correlated errors}

The errors in our $M_{X,500}$ and $Y_{500}$ measurements are correlated, since $M_{X,500}$ is used to define the integration boundary $\theta_{500}$ within which $Y_{500}$ is calculated.  However, this is a small effect since $\theta_{500} \propto M_{500}^{1/3}$.  We verified that incorporating the covariance as estimated from our $Y_{500}$ posteriors has a negligible effect on the scaling relation fit.

We also tested the importance of the effect by re-analysing the \emph{Planck} data with an updated mass prior, where the new mass is fixed to the value minimizing the offset between the original $(M_{X,500},Y_{500})$ point and the baseline scaling relation, taking into account the errors and fitted intrinsic scatter in both quantities.  The updated $Y_{500}$ values generally differ by $<5$\% from the original ones, with a few outliers with differences of up to $\approx$15\%.  Fitting the scaling relation again with original $M_{X,500}$ values and updated $Y_{500}$ values gives results entirely consistent with the original results ($<0.1\sigma$ difference in normalization and $<0.2\sigma$ difference in slope) as shown in Table~\ref{T:scaling_relation_fits} and Figure~\ref{Fi:SR_fits} and so we conclude that the correlations are negligible for the purposes of the scaling relation fit.

\subsubsection{Asymmetric mass errors}

L20 give independent upper and lower error estimates for their mass measurements, which for some clusters are significantly different.  In our baseline analysis we took the mean of the upper and lower errors as the input for \textsc{LIRA}, since it does not allow the input of asymmetric error bars.  We tested the effect of instead taking the maximum or minimum out of the asymmetric error bars as the input for \textsc{LIRA}, with negligible change to the scaling relation parameters.

\subsubsection{Malmquist bias}

\begin{figure}
\includegraphics[width=\linewidth]{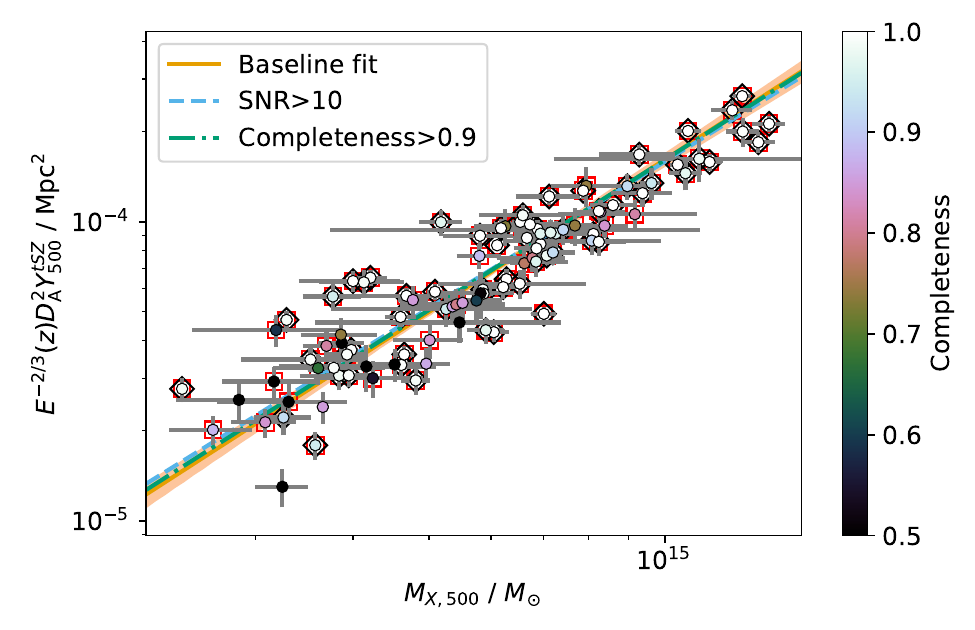}
    \caption{Testing for the presence of Malmquist bias.  The orange line and band shows the calibration of the mass-observable scaling relation using the full sample and allowing for scatter in both variables.  The blue dashed line is fitted in the same way, but restricting the sample to clusters with \textsc{PwS} SNR$>10$ (points highlighted with red squares); it is consistent within $\approx\,1\sigma$ with the full-sample fit over the whole mass range.  The dot-dashed green line shows the fit when selecting clusters with completeness $>0.9$ in the PSZ2 catalogue (points highlighted with black diamonds); it is fully consistent with the full-sample fit over the whole mass range.}
    \label{Fi:SR_malmquist}
\end{figure}

\citet{2014A&A...571A..20P} correct their individual $Y_{500}$ measurements for Malmquist bias, following \citet{2009ApJ...692.1033V} and \citet{2009A&A...498..361P}, before fitting the mean scaling relation.  However, \citet{2021ApJ...914...58A} show with simulations that the Malmquist bias is negligible for the ESZ sample when the deeper full-mission maps are used to derive SZ properties.  We tested for the presence of Malmquist bias in two ways: firstly by excluding clusters with \textsc{PwS} signal-to-noise ratio (SNR; derived from the maximum log-likelihood value, see \citealt{2012MNRAS.427.1384C}) $<10$ at which value the bias as calculated by \citet{2014A&A...571A..20P} becomes negligible (1\%).  This excludes a total of 12 out of 101 clusters of low-to-medium mass in our sample.  Fitting the scaling relation with the high-SNR sample results in only $< 0.7\sigma$ changes in the scaling relation parameters, and in the \textit{opposite} direction to that expected due to the Malmquist bias.  If $Y_{500}$ measurements at the low end of the mass range are biased high due to the Malmquist bias, the full-sample scaling relation should be biased high at low mass, however we find the opposite as shown in Figure~\ref{Fi:SR_malmquist}.

\begin{figure}
\includegraphics[width=\linewidth]{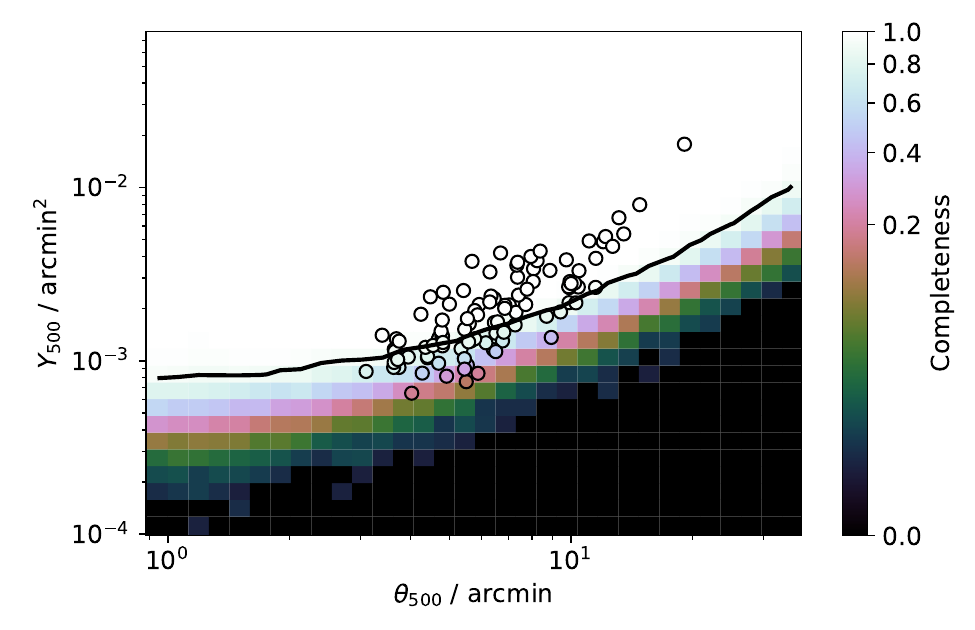}
    \caption{Completeness function for the PSZ2 catalogue (at SNR=6, used to select the cosmology sample), with the scaling relation sample shown by the open points. The black contour shows the completeness cutoff of 0.9 used to test for Malmquist bias.}
    \label{Fi:L20_completeness}
\end{figure}

Secondly, we considered the completeness expected for the cluster masses and redshift in the sample.  The PSZ2 catalogue included a completeness estimation as a function of $(\theta_{500},Y_{500})$.  For each cluster in the scaling relation sample, we estimated (a better approximation to) the true $(M_{X,500},Y_{500})$ by minimizing the offset between the measured point and the fitted scaling relation, then used this value to obtain the completeness in the PSZ2 catalogue, as illustrated in Figure~\ref{Fi:L20_completeness}.  Although we are using NPIPE data, we verified by running our analysis pipeline on the PR2 data that the SNR is highly correlated and similar ($\approx$\,10\% higher in NPIPE than PR2 on average), so the PSZ2 completeness is indicative of the NPIPE completeness.  If a cluster's true mass and redshift put it at a low completeness, this will mean that the estimated $Y_{500}$ is more likely to be biased upward for it to be included in the sample, and the SNR will also be biased upward so it may not be caught by the high-SNR selection test.  We selected a subsample with completeness $>0.9$ (68 out of 101 clusters) and fitted the scaling relation to this sample; the difference to the full-sample fit is negligible as shown in Figure~\ref{Fi:SR_malmquist}.  Based on the results of these two tests we agree with \citet{2021ApJ...914...58A} that the Malmquist bias is negligible.

\subsubsection{Redshift evolution}

We also consider allowing the redshift evolution to deviate from the self-similar expectation.  In this case, the scaling relation fit by LIRA is a little different:

\begin{equation}\label{eq:scaling_relation2}
\left [ \frac{D_\mathrm{A}^2 Y_{500}}{10^{-4} \mathrm{Mpc}^2} \right ] = Y_{*} \left [ \frac{h}{0.7} \right ]^{-2+A} \left [ \frac{M_{X,500}}{6 \times 10^{14} \mathrm{M}_{\odot}} \right ]^A \left [ \frac{E(z)}{E(z_{\mathrm{ref}})} \right ]^B,
\end{equation}
where $z_{\mathrm{ref}}=0.2$ is chosen to be near the median of the sample redshift distribution.  $Y_{*}$ in this case is therefore not directly comparable with the self-similar evolution case since it is modified by an extra factor $E(z_{\mathrm{ref}})^{-B}$.  The fit in this case is reported in Table~\ref{T:scaling_relation_fits}; the redshift evolution parameter $B$ is consistent with the self-similar value of 2/3, $B=0.3 \pm 0.7$, although the large error-bar clearly shows there is not enough redshift leverage to constrain this parameter.  We therefore restrict the redshift evolution to the self-similar case from here on.

\subsection{Monte-Carlo simulations}\label{S:selection_function}

\begin{figure}
\includegraphics[width=\linewidth]{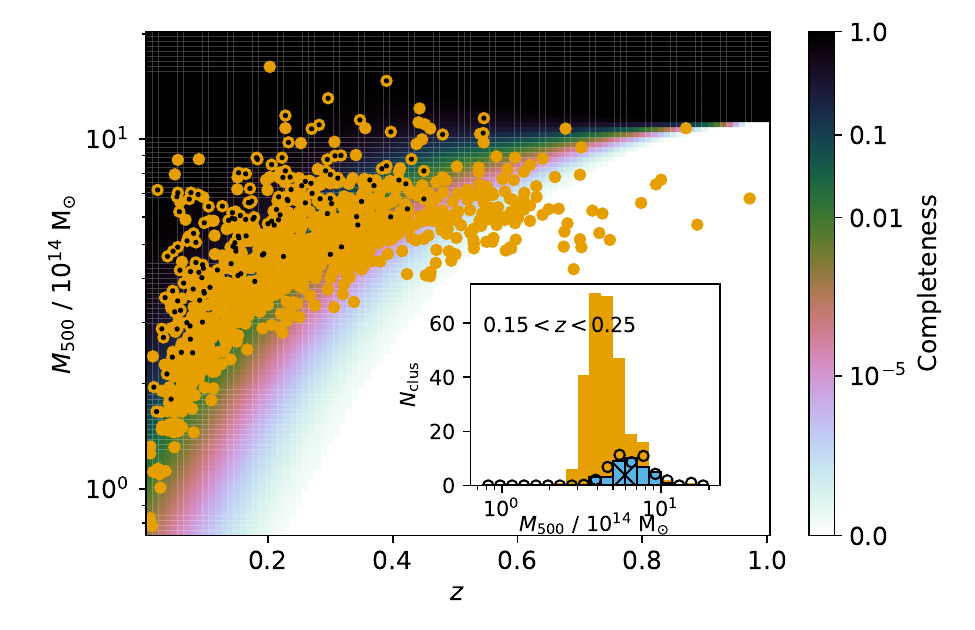}
    \caption{Completeness of the ESZ cluster sample relative to the PSZ2 sample.  The background colour-scale shows our parametric fit to the completeness.  Orange dots overplotted show the PSZ2 sample, while small black dots show clusters also present in the ESZ sample.  The inset shows the calculation for a small redshift slice.  The orange histogram shows the mass distribution of clusters in the PSZ2 in this redshift bin, while the blue hatched histogram shows the same for the overlapping ESZ and PSZ2 samples.  Open dots show the predicted numbers in each bin based on our parametric fit to the completeness.}
    \label{Fi:ESZ_completeness}
\end{figure}

We tested the accuracy of our fitting methodology and the impact of the mass-redshift selection function of our sample using Monte-Carlo simulations.  The \emph{Planck} ESZ sample did not have a well-defined selection function, however the selection function for the PSZ2 cosmological sample was robustly investigated in \citet{2016A&A...594A..24P} and publicly released in the form of the completeness function, $\chi(\theta_{500},Y_{500},\mathrm{SNR})$.  We can therefore bootstrap a completeness function for the overlapping ESZ and PSZ2 samples, i.e.\ the completeness of ESZ relative to PSZ2.  We calculated a parametric fit to the completeness as a function of mass and redshift, by fitting an error-function to the completeness as a function of mass in redshift slices, then fitting a quadratic function to the fitted error function parameters as a function of redshift.  This produces a reasonable fit (shown in Figure~\ref{Fi:ESZ_completeness} along with the real PSZ2 and ESZ detections), which we consider good enough to generate realistic simulated samples and therefore evaluate the impact of the mass-redshift selection function on the scaling relation fit.

Our Monte-Carlo simulations then proceed as follows:

\begin{enumerate}
\item Draw an expected population of clusters between the mass and redshift limits of the sample, using the \citet{2008ApJ...688..709T} mass function as implemented in the \textsc{Colossus} toolkit \citep{2018ApJS..239...35D} and apply a random 65\% sky cut to match the PSZ2 cosmological sample selection.
\item Bias the mass values by a mean hydrostatic mass bias of $(1-b)=0.7$ (empirically, this roughly matches the mass distribution to the observed PSZ2 distribution; we note we are using a concordance cosmology so we do not necessarily expect this bias to match the \emph{Planck} cosmological results).
\item Convert the mass and redshift values to $\theta_{500}$, $Y_{500}$ assuming the P13 scaling relation parameters.
\item Apply an intrinsic log-normal scatter contamination to the $Y_{500}$ values, using the mean estimate of $\sigma_{\log Y}=0.03$ from our fit to the real data.
\item Contaminate the scattered $Y_{500}$ values with random noise using the average 10\% measurement errors from our sample.
\item Filter the sample using the PSZ2 selection function, with SNR cutoff of 6, to emulate the PSZ2 cosmological sample with Malmquist bias.
\item Filter the sample again using the bootstrapped ESZ selection function to emulate the ESZ sample.
\item Contaminate mass values with intrinsic log-normal scatter ($\sigma_{\log M}=0.06$) and random noise, using the average 10\% measurement errors from our sample.
\item Discard clusters with (scattered, noisy) $\theta_{500}>30$\,arcmin to satisfy the \textit{XMM-Newton} FoV cut.
\item Fit the final contaminated values with the same \textsc{LIRA} command as we use to fit the real data.
\end{enumerate}

We repeated the simulation 500 times, fitting in three different ways:

\begin{enumerate}
\item With $B$ fixed to the input value of $2/3$, fitting for intrinsic scatter in both parameters;
\item $B$ free, fitting for scatter in both parameters;
\item $B$ fixed to the input value, no scatter fitted in $Y_{500}$.
\end{enumerate}

\begin{figure*}
\includegraphics[width=\linewidth]{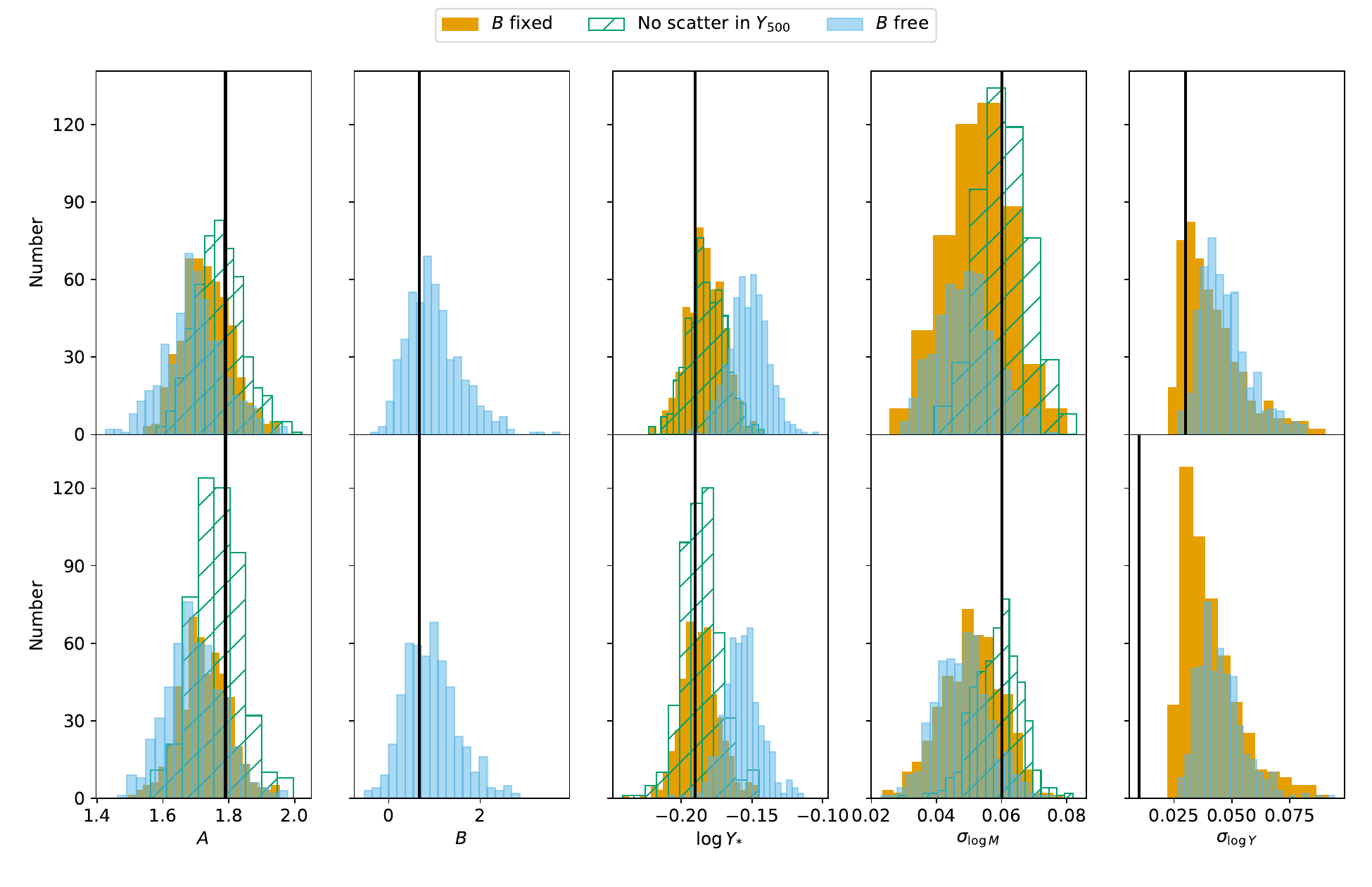}
    \caption{Results of Monte-Carlo simulations to test the robustness of the LIRA fit results given the selection effects, as described in the text.  Solid yellow histograms show results of LIRA fits to 500 random realisations of the sample, fixing $B$ to its input value of $2/3$.  Blue solid histograms show results of fits to the same realisations, also fitting $B$, and green hatched histograms show results of fits to the same realisations, not fitting for $B$ or scatter in $Y_{500}$.  Black vertical lines show input values of the scaling relation and scatter parameters.  We note that the $\log Y_{*}$ discrepancy when fitting $B$ is due to a slightly different definition (see equation~\ref{eq:scaling_relation2}) rather than an error in the fit.  In the top (bottom) row, the input intrinsic scatter in $Y_{500}$ was set to 0.03 (0.01).}
    \label{Fi:lira_tests}
\end{figure*}

We then compared all the resulting scaling relation fit parameters to check for any biases with respect to the input parameter values caused by the selection function.  Figure~\ref{Fi:lira_tests} (top row) shows the results.  The scaling relation parameters $A$, $B$ and $\log Y_{*}$ are generally fit well with all three fitting methods.  The mean fit parameters are all $<0.8\sigma$ from the true values, where $\sigma$ is the mean error reported by LIRA.  The standard deviations of the distributions are also very consistent with the errors reported by LIRA.  When $B$ is not fixed, the recovered values of $\log Y_{*}$ are inconsistent with the input value, but this is due to the slightly different definition of $\log Y_{*}$ as shown in equation~\ref{eq:scaling_relation2} rather than an error in the fit.  $B$ is constrained correctly but with very large scatter, confirming that there is not enough redshift leverage in this sample to constrain the redshift evolution.  The fit with no intrinsic scatter in $Y_{500}$ gives a slightly better result for $A$ with the mean fit parameter $<0.3\sigma$ from the true value.  There is some hint of a bias toward lower $A$ values which is probably due to the combination of the cluster mass function and the ESZ selection function, but we do not attempt to correct for this given the uncertainty in the selection function.

\begin{figure}
\includegraphics[width=\linewidth]{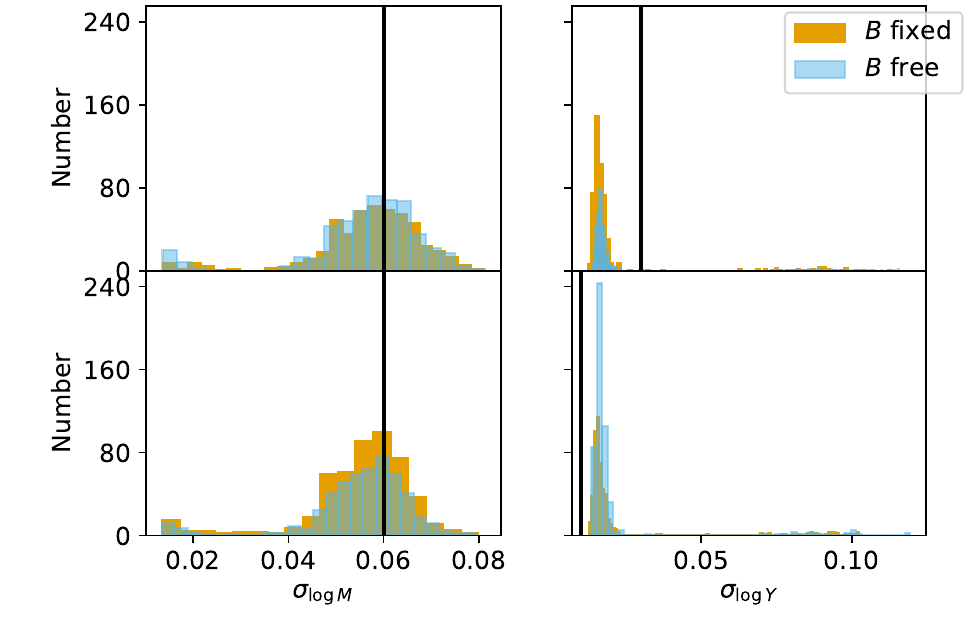}
    \caption{Results of Monte-Carlo simulations to test the robustness of the LIRA fit results given the selection effects, as described in the text.  Colours, markers and rows are as in Figure~\ref{Fi:lira_tests} except that these plots display the maximum a-posteriori values for each simulation rather than the posterior means.}
    \label{Fi:lira_tests2}
\end{figure}

The performance of the intrinsic scatter fitting is more difficult to test given the degeneracies in the posteriors of these parameters, as discussed in Section~\ref{S:intrinsic_scatter}.  Figure~\ref{Fi:lira_tests} (top row) shows the distribution of the mean values of the posteriors, in the case where $\sigma_{\log Y}=0.03$ as fit to the real data.  Since there are indications that the posterior may really be showing an upper limit, we also tested a much lower input value of $\sigma_{\log Y}=0.01$ (bottom row).  In both cases, when fitting with scatter in both observables, the mean values of $\sigma_{\log M}$ tend to be underestimated by $\approx 1\sigma$, and the mean values of $\sigma_{\log Y}$ are consistently overestimated with a very non-Gaussian distribution.  We also tested the robustness of the maximum a-posteriori (MAP) values rather than the means (see Figure~\ref{Fi:lira_tests2}).  There are a small number of cases where the MAP position occurs at the wrong end of the `elbow' shaped posterior shown in Fig.\ref{Fi:intrinsic_scatter}, i.e.\ the MAP position is at a high value of $\sigma_{\log Y}$ and a low value of $\sigma_{\log M}$.  Excluding those, the MAP values recover the true value of $\sigma_{\log M}$ well.  On the other hand, the MAP values of $\sigma_{\log Y}$ are consistently at the bottom of the prior range irrespective of the input value of 0.03 or 0.01, as in the case of the real data, confirming that intrinsic scatter in $Y_{500}$ is not constrained.  Where the posterior indicates an upper limit in $\sigma_{\log Y}$, the true value is below the 68\% upper limit 97\% (100\%) of the time in the 0.03 (0.01) case, so the upper limit derived from the posterior is conservative.

Given these results, we choose to take the fit with $B$ fixed and no scatter in $Y_{500}$ as our fiducial result.  In both the higher-scatter and lower-scatter cases, the recovered value of $A$ is less biased than when fitting with scatter ($<0.25\sigma$ mean offset from true value rather than $<0.7\sigma$), and the $\log Y_{*}$ recovery accuracy is comparable ($<0.6\sigma$ for the high-scatter and $<0.2\sigma$ for the low-scatter case).  The scatter in mass is also accurately constrained using the posterior mean and the difficulty with the correlated posterior for the two scatters is removed.

\subsection{Relativistic calibration}\label{S:relativistic_SR}

\subsubsection{Impact of temperature priors}

We carry out separate analyses using each of the \Tsz\ priors in Table~\ref{T:priors}.  We centre the Gaussian prior on the X-ray core-excised temperature measurement from L20.  For the scaling relation prior, we use the mass-\Tsz\ scaling relations derived by \citet{2020MNRAS.493.3274L} and \citet{2022MNRAS.517.5303L} based on numerical simulations:

\begin{equation}\label{eq:M_T_SR}
E(z)^{-2/3} \Tszm = A \left(\frac{M_{500}}{M_{\mathrm{fid}}} \right )^{B+C \log_x(M_{500}/M_{\mathrm{fid}})} \mathrm{keV}.
\end{equation}

We consider two different options for the scaling relations:

\begin{enumerate}
  \item The $M_{500}-T_{y,500}$ relationship derived in \citet{2020MNRAS.493.3274L} based on the BAHAMAS \citep{2017MNRAS.465.2936M} and MACSIS \citep{2017MNRAS.465..213B} hydrodynamical simulations.  In this case, $\log_x$ is the natural logarithm and $M_{\mathrm{fid}}=3\times 10^{14} h^{-1} \mathrm{M}_{\odot}$.  The fit parameters $(A, B, C)$ are given for three different redshifts in Table~\ref{T:M_T_SR}; we interpolate between them to calculate the relationship at a given cluster redshift (there are no redshifts $>1$ in the L20 or PSZ2 samples).
  \item The $M_{200}-T_{y,200}$ relationship derived in \citet{2022MNRAS.517.5303L}, averaged over results from the BAHAMAS and MACSIS, The Three Hundred Project, Magneticum Pathfinder and IllustrisTNG.  In this case, $\log_x=\log_{10}$ and $M_{\mathrm{fid}}=1\times 10^{14} \mathrm{M}_{\odot}$.  The parameters $(A, B, C) = (1.426, 0.566, 0.024)$ at $z=0$ and the right hand side is multiplied by an extra factor $A_*$, where $\log_{10}(A_*) = -0.05 \log_{10}(1+z) - 0.11 \left [\log_{10}(1+z) \right ]^2$.  This accounts for the departure from self-similar redshift evolution.
\end{enumerate}

For both the Gaussian prior and the scaling relation-based prior, rather than use an error or scatter estimate to define the prior width we simply use 2\,keV.  This is larger than the typical X-ray temperature error ($\approx 0.4$\,keV in L20) and intrinsic variance in the numerical simulations (7 per cent for $T_y$ quoted in \citealt{2022MNRAS.517.5303L}), to reflect the fact that there are uncertainties in the difference between X-ray and SZ temperatures and in the gas physics used in the numerical simulations from which the scaling relations are defined. 2\,keV is an arbitrary, conservative choice, however we note that the error in $Y_{500}$ is driven more by the profile parameter shape uncertainty than the temperature error.  For a high-significance cluster with temperature $\approx$\,10\,keV, decreasing the prior width (with the scaling relation-based prior) from 2\,keV to 1\,keV results in a negligible change in the mean $Y_{500}$ estimate of 0.4\% and a negligible change in $\Delta Y_{500}/Y_{500}$ from 6.7\% to 6.6\%.

\begin{table}
	\centering
	\caption{Fitting parameters given in \citet{2020MNRAS.493.3274L} for the scaling relation between $M_{500}$ and \Tsz\ as defined in equation~\ref{eq:M_T_SR}.}
	\label{T:M_T_SR}
	\begin{tabular}{lccc}
		\hline
		$z$ & $A$ & $B$ & $C$ \\
		\hline
                0.0 & 4.763 & 0.581 & 0.013 \\
                0.5 & 4.353 & 0.571 & 0.008 \\
                1.0 & 3.997 & 0.593 & 0.009 \\
		\hline
	\end{tabular}
\end{table}

To implement the $M_{200}-T_{y,200}$ scaling relation, we need to convert from the X-ray $M_{X,500}$ values to an $M_{200}$ estimate.  We do this using the concentration-mass relation calibrated in \citet{2021MNRAS.506.4210I} for $r_{500}$, combined with the conversion formula from \citet{2003ApJ...584..702H} (their Appendix C) which allows us to calculate $r_{200}$ and hence $M_{200}$.  The $M_{X,500}/M_{200}$ ratios from this calculation are consistent with the simulation-based average $M_{500}/M_{200}$ ratios in Table B5 of \citet{2022MNRAS.517.5303L}, giving confidence in the calculation.

\begin{figure}
\includegraphics[width=\linewidth]{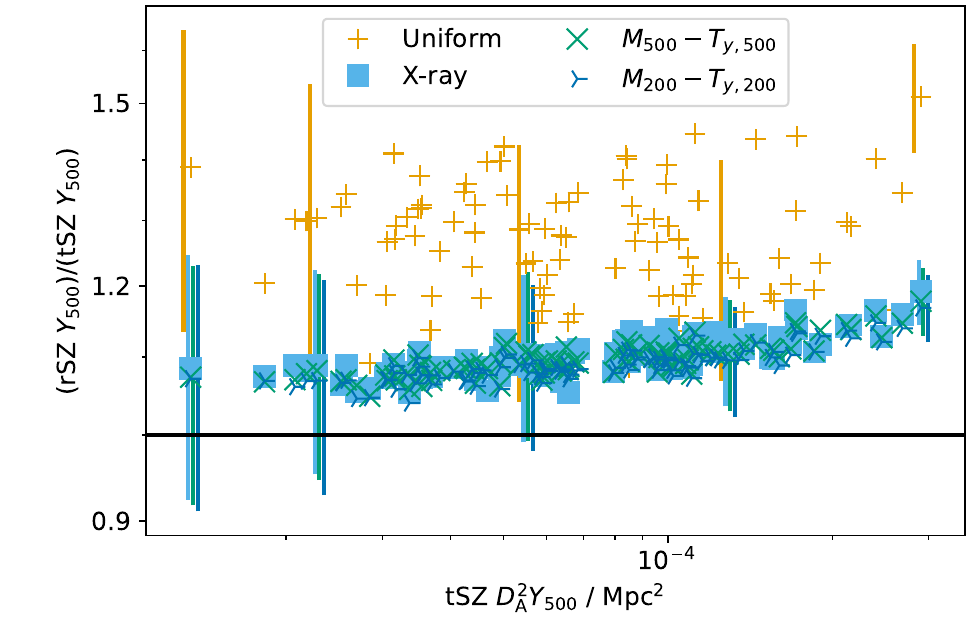}
    \caption{Comparison between $Y_{500}$ constraints derived using the non-relativistic SZ spectrum and the relativistic spectrum, with our three different priors (and two different scaling relations) on \Tsz.  All results use X-ray priors on $M_{X,500}$.  Some representative errorbars are shown (transposed slightly horizontally for clarity) and the black horizontal line marks a ratio of $1$.} It is clear that the uniform prior results in unfeasibly large boosts in $Y_{500}$ as expected from simulations, while the results using an X-ray measurement of temperature as a \Tsz\ prior are very similar to those using the mass-temperature scaling relations.
    \label{Fi:Xray_cal_comp}
\end{figure}

Results are reported in Table~\ref{T:Y500_Xray_cal}.  In Figure~\ref{Fi:Xray_cal_comp} we compare the $D_\mathrm{A}^2 Y_{500}$ values obtained using the three different priors (and two different scaling relations) on \Tsz\ to the non-relativistic results.  As expected from the simulation results, the $Y_{500}$ constraints obtained using a uniform prior on \Tsz\ are biased upward by unrealistically large amounts for most clusters, and so we do not use these results any further.  The $Y_{500}$ constraints obtained using the X-ray temperature measurement as a prior on \Tsz\ are very similar to those obtained using the mass-\Tsz\ scaling relations.

\begin{table*}
\centering
\caption{$Y_{500}$ constraints from fitting \emph{Planck} data using X-ray priors on $M_{X,500}$ from L20 as a prior constraint on $\theta_{500}$ and varying the profile shape parameters.  $z$ is redshift; $M_{X,500}$ and $\Delta M_{X,500}$ are the L20 mass measurement and errorbar (the average of the upper and lower limits); $k T_{\mathrm{exc}}$ is the L20 core-excised X-ray temperature measurement.  The $D_\mathrm{A}^2 Y_{500}$ measurements and corresponding errorbars are, from left to right: tSZ measurement; rSZ with $M_{200}-T_{y,200}$ temperature prior; rSZ with X-ray temperature prior centred on $k T_{\mathrm{exc}}$.  The first ten rows of the table are shown here; the full table is available as supplementary material.}
\label{T:Y500_Xray_cal}
\footnotesize
\begin{tabular}{lcccccccccc}
\hline
PSZ2 & $z$ & $M_{X,500}$ & $\Delta M_{X,500}$ & $k T_{\mathrm{exc}}$ & $D_A^2 Y_{500,\mathrm{tSZ}}$ & $\Delta D_A^2 Y_{500,\mathrm{tSZ}}$ & $D_A^2 Y_{500,\mathrm{M-}\Tszm}$ & $\Delta D_A^2 Y_{500,\mathrm{M-}\Tszm}$ & $D_A^2 Y_{500,\mathrm{T_{X}}}$ & $\Delta D_A^2 Y_{500,\mathrm{T_{X}}}$ \\
& & \multicolumn{2}{c}{/ $10^{14}$ M$_{\odot}$ } & / keV & \multicolumn{6}{c}{/ $10^{-5}$ Mpc$^2$} \\
\hline
G000.40-41.86 & 0.165 & 5.01 & 0.52 & 5.82 & 4.24 & 0.55 & 4.53 & 0.62 & 4.61 & 0.61 \\
G002.77-56.16 & 0.141 & 4.96 & 0.35 & 5.39 & 3.49 & 0.44 & 3.83 & 0.49 & 3.82 & 0.46 \\
G003.93-59.41 & 0.151 & 6.94 & 0.19 & 6.46 & 8.81 & 0.59 & 9.69 & 0.65 & 9.79 & 0.67 \\
G006.68-35.55 & 0.089 & 2.42 & 0.04 & 4.62 & 2.85 & 0.24 & 2.98 & 0.26 & 3.01 & 0.26 \\
G008.47-56.34 & 0.149 & 3.61 & 0.08 & 4.91 & 3.37 & 0.42 & 3.50 & 0.42 & 3.55 & 0.44 \\
G008.94-81.22 & 0.307 & 10.39 & 0.23 & 8.43 & 17.23 & 1.11 & 19.51 & 1.38 & 19.61 & 1.34 \\
G021.10+33.24 & 0.151 & 6.88 & 0.19 & 8.77 & 9.96 & 0.72 & 10.92 & 0.80 & 11.33 & 0.87 \\
G039.85-39.96 & 0.176 & 3.77 & 0.32 & 5.82 & 5.88 & 0.62 & 6.26 & 0.68 & 6.39 & 0.66 \\
G042.81+56.61 & 0.072 & 4.65 & 0.14 & 4.79 & 3.68 & 0.25 & 3.96 & 0.28 & 3.95 & 0.29 \\
G046.10+27.18 & 0.389 & 6.26 & 0.54 & 5.65 & 10.90 & 1.60 & 11.83 & 1.71 & 11.83 & 1.76 \\
\hline
\end{tabular}

\end{table*}

\begin{figure}
\includegraphics[width=\linewidth]{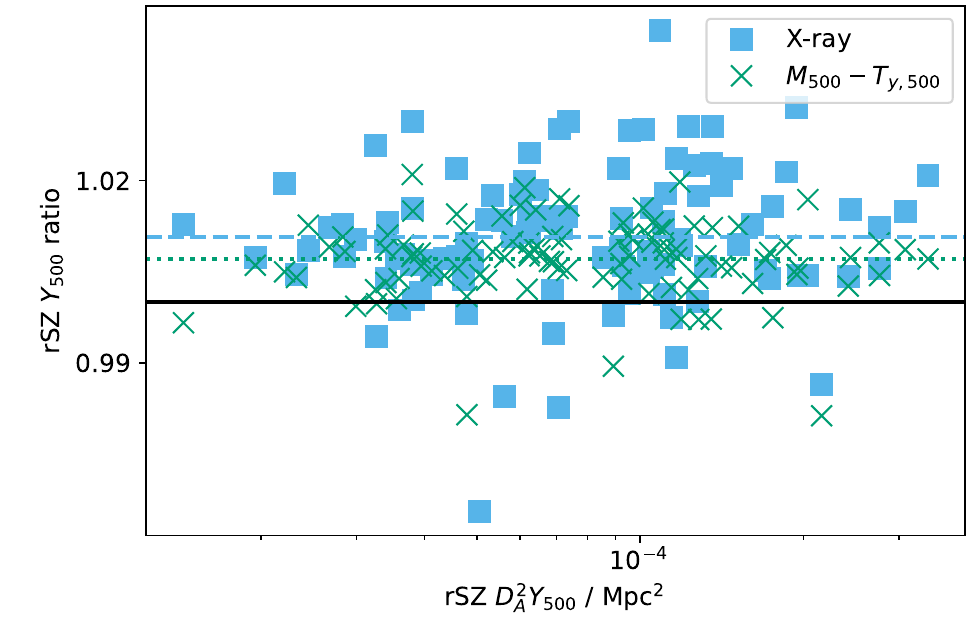}
    \caption{Comparison between relativistic $Y_{500}$ constraints derived using the $M_{200}-T_{y,200}$ scaling relation ($x$-axis) and the X-ray and $M_{500}-T_{y,500}$ scaling relation prior (ratio on $y$-axis).  The blue dashed and green dotted horizontal lines show the mean ratios, which are very close to one (the black solid line).}
    \label{Fi:temp_comparisons}
\end{figure}

In Section~\ref{S:temp_systematics} we showed that \Tsz\ decreased when averaged over larger cluster volumes, and that the X-ray temperature (averaged within $r_{500}$) could be significantly larger than the globally-averaged \Tsz.  The higher temperatures could therefore produce a systematic bias in the $Y_{500}$ values.  We investigate this issue in Figure~\ref{Fi:temp_comparisons} further by comparing the different constraints produced by the X-ray temperature and $M_{500}-T_{y,500}$ scaling relations to those produced by the $M_{200}-T_{y,200}$ scaling relation, which should be closest to the globally-averaged temperature.  We see that the $Y_{500}$ constraints are biased very slightly higher with the higher temperatures, but only by a mean of 1 per cent in the case of the X-ray temperature and 0.7 per cent in the case of the $M_{500}-T_{y,500}$ scaling relation, with no sign of a mass-dependent bias.  This gives us confidence that although we do not have a measurement of the globally-averaged \Tsz\ currently, as the temperature measurements converge to a closer approximation to it we obtain $Y_{500}$ measurements that converge within a very small offset.  Our results using the $M_{200}-T_{y,200}$ scaling relation should therefore be very close to unbiased.

\subsubsection{Relativistic $M_{X,500}-Y_{500}$ scaling relation}

\begin{figure}
\includegraphics[width=\linewidth]{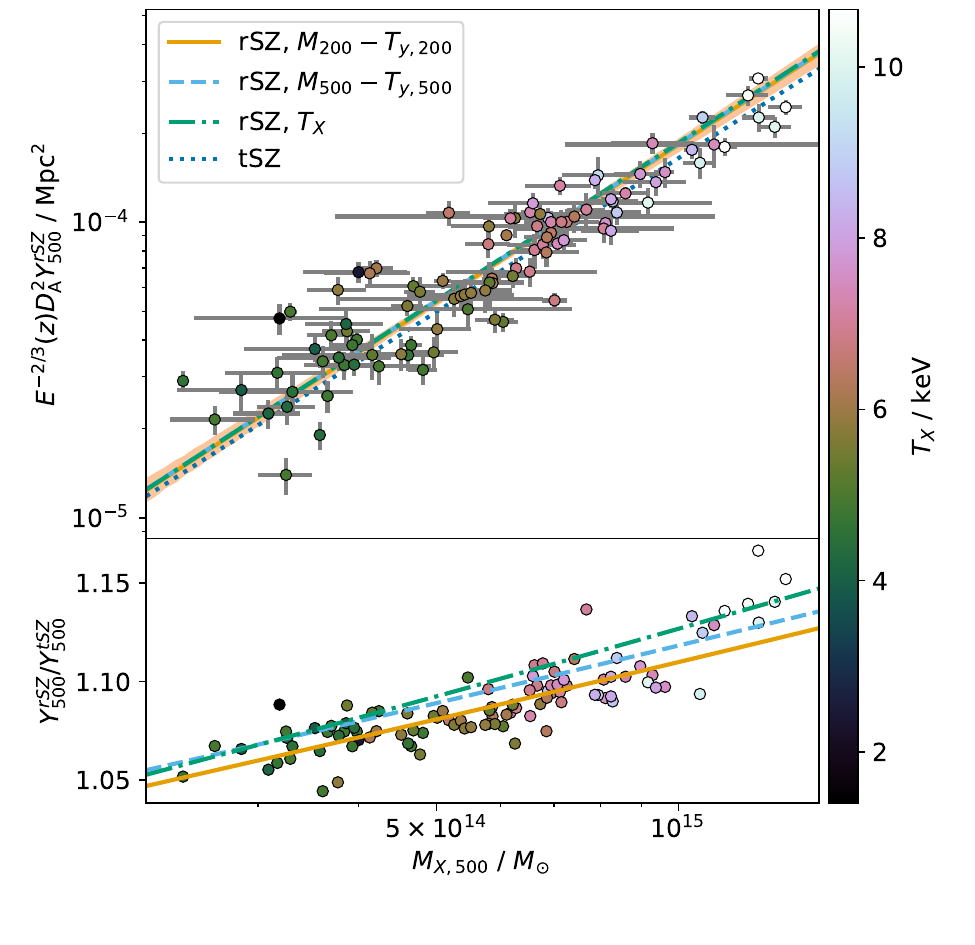}
    \caption{Top panel: calibration of the mass-observable scaling relation assuming the relativistic SZ spectrum and $M_{200}-T_{y,200}$ scaling relation on temperature (orange line, shaded area showing uncertainty) in comparison with the other temperature priors and the non-relativistic calibration (dark blue dotted line).  The bottom panel shows the ratio between the relativistic and non-relativistic results, with the points showing the estimates derived with the $M_{200}-T_{y,200}$ prior.}
    \label{Fi:SR_fits2}
\end{figure}

Figure~\ref{Fi:SR_fits2} shows the scaling relation fit when the relativistic SZ spectrum is applied, comparing all three of the informative priors on \Tsz.  We use the fiducial method with no scatter in $Y_{500}$ to perform the fit.  As expected from the simulations, there is a mass-dependent change in the calibration rising to $\approx$12\% at the high-mass end of the sample.  We note that this does not actually imply that \emph{Planck} cluster masses are underestimated; in analyzing \emph{Planck} data, since the biased $M_{X,500}-Y_{500}$ calibration is applied to similarly biased $Y_{500}$ measurements of other clusters, the output mass will not be biased.  However, if the scaling relation calibrated from \emph{Planck} data is applied to samples of clusters measured by other instruments with different frequency bands, the output mass may be biased.

At the high-mass end of the sample ($M_{X,500}>10^{15} M_\odot$), the ratio between the relativistic and non-relativistic $Y_{500}$ estimates is somewhat higher than expected based on the ratio between the relativistic and non-relativistic scaling relation \emph{fits}.  This may indicate that there is a more complicated relationship between temperature and mass for these high-mass clusters than is indicated by our assumed scaling relations.  Further investigation of this will require temperature measurements in the outskirts of these clusters.

\section{Discussion}

\subsection{rSZ effect}

It is clear that relativistic effects, often treated as negligible in SZ literature in the past, must now be taken into account given the sensitivity of current instruments.  Assuming the non-relativistic thermal approximation to the SZ spectrum results in a bias in Compton-$y$ estimates which are used to derive cluster mass estimates and extract pressure profiles.  When scaling relations are calibrated from and applied to SZ observations from the same instrument, this does \emph{not} result in a mass bias.  However since the bias is frequency-dependent, when a scaling relation calibrated from one SZ instrument is applied to observations by another instrument in different frequency bands (e.g.\ SZ observations with NIKA2 at 150 and 260\,GHz compared to \emph{Planck} scaling relations in \citealt{2024A&A...684A..18A}) or compared to numerical simulations (e.g.\ \citealt{2017MNRAS.469.3069G}) it is important to use unbiased scaling relations.  Similarly, when information from different SZ instruments observing in different bands is combined (for example to create combined $y$-maps, e.g.\ \citealt{2022ApJS..258...36B}, or fit combined scaling relations, e.g.\ \citealt{2022ApJ...934..129S}), the relativistic spectrum should be considered to avoid systematic differences between instruments.  Pressure profile constraints, particularly for massive clusters, will be biased if relativistic corrections are not accounted for; this is starting to be done in the literature, for example \citet{2023ApJ...944..221S} applies relativistic corrections to their SZ data based on X-ray temperature profiles.  Although the correction is small, in the quest for precise cluster mass estimates that can be used for precision cosmology it is non-negligible.

With current instruments, the rSZ temperature is just at the limit of detectability/constraint.  This is shown here for \emph{Planck}.  Other attempts to measure the rSZ spectrum on individual clusters include \citet{2010A&A...518L..16Z} using \emph{Herschel-SPIRE} data, and \citep{2022ApJ...932...55B} using \emph{Herschel-SPIRE}, \emph{Bolocam}, and \emph{Planck} data.  Both of these studies achieve low-significance detections of the rSZ spectrum, and only \citet{2022ApJ...932...55B} independently measures an rSZ temperature, with large errorbars, in common with the results presented here.  Future instruments such as the Fred Young Submillimeter Telescope \citep{2023ApJS..264....7C} and the Atacama Large Aperture Submillimetre Telescope \citep{2022SPIE12190E..07R} which have the high-frequency observing bands necessary to constrain the rSZ corrections, along with improved angular resolution and sensitivity compared to current instruments will open up new possibilities in this area.

\subsection{Updated scaling relation}

\begin{figure}
\includegraphics[width=\linewidth]{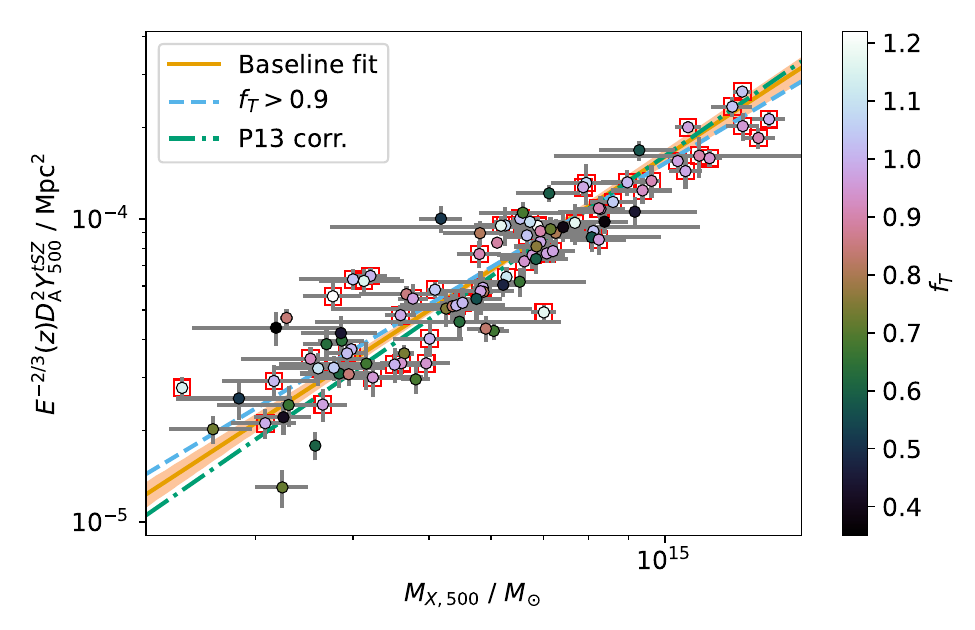}
    \caption{Testing the effect of incomplete radial coverage in X-ray.  The parameter $f_T$ shown in the colour-scale represents the fraction of $r_{500}$ that is covered by the X-ray temperature profile.  The orange line and band shows the calibration of the mass-observable scaling relation using the full sample and allowing for scatter in both variables.  The blue dashed line is fitted in the same way, but restricting the sample to clusters with $f_T>0.9$ (points highlighted with red squares).  The dot-dashed green line shows the P13 fit.}
    \label{Fi:SR_fT}
\end{figure}

Apart from the rSZ correction at the massive end, another notable aspect of our updated scaling relation is the deviation from the P13 scaling relation at the low-mass end, by $\approx$\,10\% (Figure~\ref{Fi:SR_fits}).  This would imply a \emph{decrease} in the \emph{Planck} masses at $M_{X,500} \lessapprox 5 \times 10^{14}$ M$_\odot$, exacerbating the tension between cosmological parameters derived from the \emph{Planck} number counts and from the primordial CMB anisotropy \citep{2014A&A...571A..20P} and requiring an even larger mass bias to reconcile them.  The L20 sample has more clusters in the low-mass region than the P13 sample, but fitting a scaling relation to our updated $Y_{500}$ measurements of the P13 sample gives equivalent results.  We investigated possible systematics in the X-ray measurements by refitting the scaling relation with the sample restricted to clusters where the X-ray temperature measurements extended to $>0.9\times r_{500}$, as shown in Figure~\ref{Fi:SR_fT}.  This had a greater impact on the scaling relation fit than any of the other changes made previously, and would further exacerbate the mass bias problem, however also implies another selection function to be defined and investigated.

The low-mass discrepancies highlight the need for larger calibration samples covering the full mass range of the overall cluster sample such as the Cluster HEritage project with XMM-Newton project \citep{2021A&A...650A.104C}, and with measurements covering the full radial range of interest.  This will minimize systematic errors, narrow down the error in the scaling relation fit and avoid extrapolation in mass.  The potential bias in the slope parameter $A$ shown in our Monte-Carlo simulations (Figure~\ref{Fi:lira_tests}) also highlights the need for a well-understood selection function which can be accounted for in the scaling relation fitting process.  If our $A$ estimate is biased low, increasing it slightly would bring the scaling relation back more into line with the P13 relation, but we are not confident enough in the bootstrapped ESZ selection function to make this correction.

\subsection{Profile shape}

A further small correction we have made here is to allow the profile shape parameters to vary for each cluster.  Even at the resolution of \emph{Planck}, this did make a significant difference to some estimates of $Y_{500}$, particularly at the high-mass end of the scaling relation where visually the scatter decreased compared to $Y_{500}$ values extracted with a fixed UPP profile.  For the less resolved clusters, this change mostly resulted in increased error bars on $Y_{500}$, better reflecting the uncertainty in the estimate.  This may be one reason for the lack of scatter detected in $Y_{500}$ in our scaling relation fit, in contrast with previous studies.

\section{Conclusions}

We have investigated the impact of relativistic SZ corrections on \emph{Planck} measurements of massive galaxy clusters, along with the impact on pressure profile variation.  We have done so using realistic simulations, in a fully Bayesian framework.  We have tested the impact of prior assumptions, degeneracies in the posteriors, and assessed the accuracy of the recovered posteriors.  In particular, we find:

\begin{enumerate}
\item Relativistic corrections are non-negligible for \emph{Planck} cluster analysis, producing biases of $\approx$\,5 -- 15\% and up to $\approx$\,3$\sigma$ in integrated Compton-$y$ estimates when not accounted for.
\item Weak temperature constraints are possible based on \emph{Planck} data only for some of the highest signal-to-noise-ratio clusters, however most clusters require external temperature information for accurate Compton-$y$ constraints.
\item Correlated dust emission is well-handled by the analysis framework, and does not cause a bias in recovered Compton-$y$ or temperature constraints.
\item An isothermal model is accurate enough for \emph{Planck} analysis, although there are indications (based on evidence differences between analysis of simulations with resolved versus isothermal temperature models) that a resolved temperature model would be more appropriate.
\item Systematic differences in temperature measurements are important, and the best current prior information on temperature for rSZ measurements is derived from numerical simulations rather than X-ray measurements.
\item Profile parameter shape assumptions also have a non-negligible impact for Compton-$y$ constraints for high signal-to-noise-ratio clusters.  When external constraints on the mass are available from other measurements or from a scaling relation, uncertainty in the profile can be marginalized over.
\end{enumerate}

Informed by the results of this investigation, we have recalibrated the \emph{Planck} mass-observable scaling relation for galaxy clusters.  We have used the updated NPIPE data and improved methodology, alongside a larger sample of \emph{XMM-Newton} hydrostatic masses than were used in the original scaling relation fit.  We find that:

\begin{enumerate}
\item The \emph{Planck} mass-observable scaling relation can be calibrated in a robust way using $y$-weighted temperature scaling relations from simulations toaccount for the relativistic SZ effect.
\item The relativistic corrections induce an $\approx$\,12\% change in the scaling relation at the high-mass end.
\item There is a hint of deviation from the previous \emph{Planck} scaling relation at the low-mass end, which would make \emph{Planck} masses around 10\% lower.  However better understanding of the selection function is required to confirm this.
\end{enumerate}

\begin{acknowledgement}
The author thanks Richard Saunders for useful discussion on the content of this paper, and an anonymous referee for a careful review and insightful comments.  Computations were performed on the R\={a}poi high-performance computing facility of Victoria University of Wellington.
\end{acknowledgement}

\paragraph{Funding Statement}

YCP is supported by a Rutherford Discovery Fellowship, managed by Royal Society Te Ap\={a}rangi.

\paragraph{Competing Interests}

None.

\paragraph{Data Availability Statement}

\emph{Planck} data are publicly available from the \emph{Planck} Legacy Archive: \url{https://pla.esac.esa.int/}.  Analysis code may be made available upon reasonable request to the author.

%\endnote in some journals will behave like \footnote; and \printendnotes will not output anything. 
\printendnotes

\bibliography{planck_rsz}

\end{document}